\documentclass[aps,prd,amsmath,amssymb,preprintnumbers,groupedaddress,twocolumn,nofootinbib]{revtex4-1}
\usepackage{tikz}
\usepackage{slashed}
\usepackage{hyperref}
\usepackage{verbatim}
\usepackage[normalem]{ulem}

\newcommand{\lsim}{\!\mathrel{\hbox{\rlap{\lower.55ex \hbox{$\sim$}} \kern-.34em \raise.4ex \hbox{$<$}}}}
\newcommand{\gsim}{\!\mathrel{\hbox{\rlap{\lower.55ex \hbox{$\sim$}} \kern-.34em \raise.4ex \hbox{$>$}}}}
\newcommand{\hc}{\text{ h.c.}}
\newcommand{\vev}[1]{ \left\langle {#1} \right\rangle }
\newcommand{\abs}[1]{ \left| {#1} \right| }
\def\be{\begin{equation}}
\def\ee{\end{equation}}
\newcommand{\Eref}[1]{Eq.~(\ref{#1})}
\newcommand{\Erefs}[2]{Eqs.~(\ref{#1}) and~(\ref{#2})}
\newcommand{\Fref}[1]{Fig.~\ref{#1}}
\newcommand{\Sref}[1]{Sec.~\ref{#1}}
\newcommand{\Aref}[1]{App.~\ref{#1}}

\newcommand{\Rref}[1]{Ref.~\cite{#1}}
\newcommand{\Rrefs}[1]{Refs.~\cite{#1}}
\def\ie{{\it i.e.}}
\def\eg{{\it e.g.}}
\def\GeV{\text{ GeV}}
\def\TeV{\text{ TeV}}

\def\Mfive{{MCHM$_{5+1}$}}
\def\betaexp{\ensuremath{\beta_{\rm SM}}}

\def\beq{\begin{equation}}
\def\eeq{\end{equation}}

\begin{document}
\setlength{\baselineskip}{0.22in}

\preprint{FERMILAB-PUB-16-076-T}
\title{Tadpole-Induced Electroweak Symmetry Breaking and pNGB Higgs Models}
\author{Roni Harnik, Kiel Howe, John Kearney} \affiliation{Theoretical Physics Department, Fermi National Accelerator Laboratory\\ Batavia, IL 60510 USA}
\date{\today}

\begin{abstract}
We investigate induced electroweak symmetry breaking (EWSB) in models in which the Higgs is a pseudo-Nambu-Goldstone boson (pNGB).
In pNGB Higgs models, Higgs properties and precision electroweak measurements imply a hierarchy between the EWSB and global symmetry-breaking scales, $v_H \ll f_H$. When the pNGB potential is generated radiatively, this hierarchy requires fine-tuning to a degree of at least $\sim v_H^2/f_H^2$.
We show that if Higgs EWSB is induced by a tadpole arising from an auxiliary sector at scale $f_\Sigma \ll v_H$, this tuning is significantly ameliorated or can even be removed.
We present explicit examples both in Twin Higgs models and in Composite Higgs models based on $SO(5)/SO(4)$. For the Twin case, the result is a fully natural model with $f_H \sim 1$~TeV and the lightest colored top partners at 2 TeV. These models also have an appealing mechanism to generate the scales of the auxiliary sector and Higgs EWSB directly from the scale $f_H$, with a natural hierarchy $f_\Sigma \ll v_H \ll f_H \sim{\rm TeV}$. The framework predicts modified Higgs coupling as well as new Higgs and vector states at LHC13.
\end{abstract}

\maketitle

\section{Introduction}
\label{sec:intro}

The discovery of the Higgs boson has sharpened the problem of the naturalness of the electroweak (EW) scale. An attractive solution is that the Higgs boson is a composite pseudo-Nambu-Goldstone Boson (pNGB) of a global symmetry that is spontaneously broken at a scale $f_H$ not far above the electroweak scale $v_H=246\GeV$~\cite{Kaplan:1983fs,Kaplan:1983sm}. More modern realizations of this idea include Composite Higgs (CH) models (with partial compositeness)~\cite{Kaplan:1991dc,Contino:2004vy,Agashe:2004rs}, as well as Twin Higgs (TH)~\cite{Chacko:2005pe,Chacko:2005un} and Little Higgs~\cite{ArkaniHamed:2002qx,ArkaniHamed:2002qy,Low:2002ws,Kaplan:2003uc}.

Standard Model (SM) interactions must explicitly break the global symmetries protecting the pNGB Higgs. This results in radiative contributions to the pNGB potential, with the largest contributions from the top Yukawa coupling and the gauging of $SU(2)_L$.
These contributions connect the mass scales of new top and gauge partners restoring the global symmetries to the mass scale of the Higgs boson, and in minimal composite Higgs models the pNGB potential is \emph{entirely} generated by these contributions.
For instance, the contributions from the top sector perturb the vev and physical Higgs mass proportionally by an amount of size
\be
\label{eq:radtoptuning}
\abs{\delta m_h^2} \gsim \frac{3 y_t^2}{4 \pi^2} m_\ast^2 \sim (125\GeV)^2 \left(\frac{m_\ast}{500 \GeV}\right)^2
\ee
where $m_\ast$ is the mass scale of the top partners which restore the global symmetry. If these resonances are sufficiently light, the physical Higgs mass $m_h=125\GeV$ can be obtained naturally without any tuning. Direct experimental limits on the scale $m_\ast$ of top partners \cite{Aad:2015kqa,Khachatryan:2015oba,Aad:2016qpo} give lower bounds on the tuning of such theories, but current bounds can allow a totally natural mass scale for the Higgs when colored top partner decays are hidden \cite{Anandakrishnan:2015yfa,Serra:2015xfa} or the global symmetry is partially restored by neutral particles, as in Twin Higgs models \cite{Chacko:2005pe,Chacko:2005un}.

However, observations of Higgs properties \cite{ATLAS-CONF-2015-044,Khachatryan:2014jba,Aad:2015gba} require $v_H \ll f_H$ so that the curvature of the pNGB manifold does not induce significant Higgs coupling deviations from the SM values (see, \eg, \cite{Bellazzini:2014yua,Panico:2015jxa}). SM-like Higgs measurements at the level of $\sim 10\%$ constrain $\frac{f_H^2}{v_H^2} \gsim 10$, and future measurements will reach the $\sim1\%$ level \cite{Dawson:2013bba,Asner:2013psa,Peskin:2013xra}.
This makes realizing a natural model much more difficult.
Minimal versions of 3rd generation partners can only obtain $m_h=125\GeV$ when $v_H \ll f_H$ with severe radiative tuning~\cite{Bellazzini:2014yua,Panico:2015jxa}.
More elaborate/extended fermionic sectors can improve the situation, but the structure of radiative contributions to the pNGB potential still leads to an `irreducible' tuning $\Delta \gsim \frac{f_H^2}{2 v_H^2}$.

These obstacles motivate studying pNGB Higgs models with a combination of additional tree-level contributions to the potential and top sectors that minimize radiative contributions, as such models stand the best chance to be `maximally natural.'
One well-known strategy, used in Little Higgs (as well as some TH models~\cite{Chacko:2005vw}), is to introduce additional dynamics generating a tree-level quartic without a significant contribution to the Higgs mass-squared parameter. The quartic is the dominant term in the potential, stabilizing the vacuum at $v_H=0$, and the radiative potential, which generates a negative mass-squared parameter, is a small perturbation moving the vacuum to a non-zero vev with a natural hierarchy $v_H \ll f_H$.

\begin{figure*}
\centering
Radiative EWSB:\qquad\qquad\qquad\qquad\qquad\qquad\qquad Induced EWSB: \qquad\qquad\qquad
\includegraphics[width=0.75\textwidth]{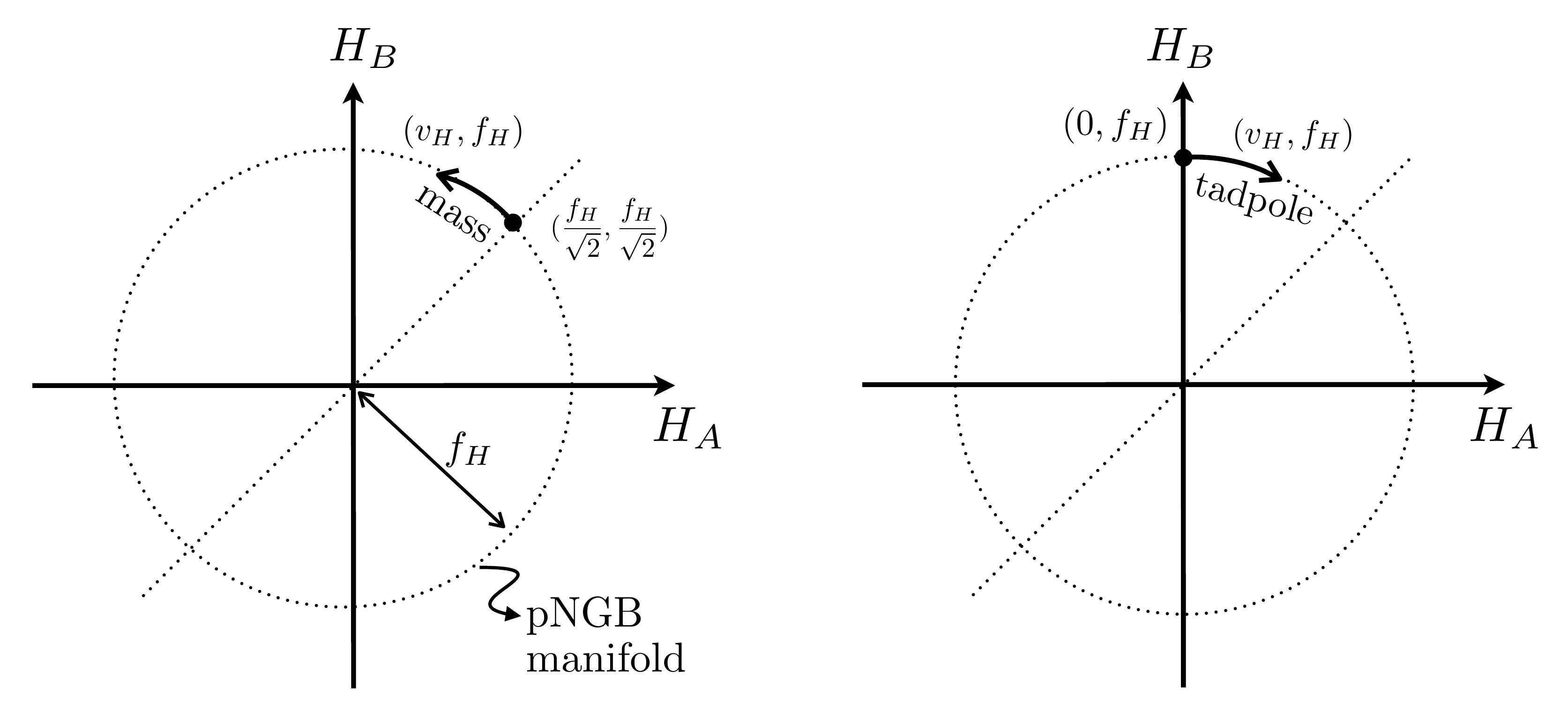}
\caption{
\label{fig:cartoon}
A schematic depiction of ``regular'' radiative EWSB (left) versus induced EWSB (right) in a pNGB Higgs model. In this figure we take the Twin Higgs as an example where $H_A$ is the SM Higgs doublet and $H_B$ is its mirror partner (but the mechanism applies more broadly). In both cases the non-linear sigma model constrains the vev to live on a ``pNGB manifold'' (dotted circle).
In the radiative EWSB the generic, untuned, EW vev is tuned down from $f_H$ to $v_H$ using a mass term. In the induced case an untuned EW vev of zero is brought up to $v_H$ without tuning by a tadpole.}
\end{figure*}

Here, we study an alternative approach. 
The pNGB potential will naturally be of the size of the radiative contributions, but with a positive mass-squared stabilizing the vacuum at $v_H = 0$.
An auxiliary decoupling EWSB sector $\Sigma$ is then introduced to trigger Higgs EWSB through a linear coupling to the Higgs sector, perturbing the Higgs vacuum to a non-zero vev with a natural hierarchy $f_\Sigma \ll v_H \ll f_H$ (where the total scale of EWSB is $v^2=f_\Sigma^2 + v_H^2$).  This is an application of Bosonic Technicolor (BTC) or, as it is more recently dubbed, induced EWSB
~\cite{Simmons:1988fu,Samuel:1990dq,Dine:1990jd,Kagan:1990az,Kagan:1991gh,Kagan:1992aq,Carone:1992rh,Carone:1993xc,Antola:2009wq,Antola:2011bx,kagantalk,Azatov:2011ht,Azatov:2011ps,Galloway:2013dma,Chang:2014ida} to a pNGB Higgs. A schematic comparison of this approach to the tuned minimal radiative approach is shown in \Fref{fig:cartoon}. 

The tuning problem in pNGB models in many ways resembles the little hierarchy problem of the minimal supersymmetric standard model (MSSM), where obtaining $m_h=125\GeV$ radiatively requires stop masses $m_{\tilde t} \gg$ TeV and/or large A-terms, both of which directly contribute to the tuning~\cite{Barbieri:2000gf}. It is not surprising then that parallels can be drawn between proposed solutions in the two frameworks. For example, the addition of radiatively-safe tree-level quartics is commonplace both in supersymmetric models (as in, \eg, the NMSSM)~\cite{Batra:2003nj, Harnik:2003rs, Hall:2011aa, Bertuzzo:2014sma} and in Little Higgs. Indeed, the approach we take here to reconciling the Higgs mass with naturalness has been considered previously in the context of supersymmetric models 
\cite{Samuel:1990dq,Dine:1990jd,kagantalk,Azatov:2011ht,Azatov:2011ps,Galloway:2013dma,Chang:2014ida}, 
but has not yet been employed in composite/pNGB Higgs models.  

In the subsequent sections, we will explore the details of tadpole-induced EWSB in two concrete realizations of composite Higgs models---`conventional' Minimal Composite Higgs models (MCHM) based on $SO(5)/SO(4)$~\cite{Agashe:2004rs,Bellazzini:2014yua,Panico:2015jxa} and composite Twin Higgs models~\cite{Chacko:2005pe,Chacko:2005un,Barbieri:2015lqa,Low:2015nqa} based on $SO(8)/SO(7)$  (or $SU(4)/SU(3)$ for weakly-coupled UV completions). The ability of tadpole-induced EWSB to improve naturalness differs for the two frameworks: \\\\
\underline{$SO(5)/SO(4)$ MCHM:}
For $SO(5)/SO(4)$ models, tadpole-induced EWSB can easily allow $m_h=125\GeV$.
This makes models with minimal breaking of the global symmetry and minimal representations of fermionic third generation partners (\Mfive) viable. These models normally fail to generate a sufficiently heavy Higgs without excessive tuning.
However, the tuning in these models still remains larger  than $\frac{f_H^2}{2 v_H^2}$ even with induced EWSB due to size of radiative contributions from the top sector, and so the tadpole mechanism does not necessarily improve naturalness compared to models that obtain $m_h=125\GeV$ via radiative contributions from an extended top sector (\eg, ${\rm MCHM}_{14+1}$ with very light top partners) or large $\tau_R$ compositeness.
So tadpole-induced EWSB resuscitates some (in particular, simple) composite Higgs models, but does not necessarily improve upon more complicated, minimally-tuned models.\\\\
\underline{Twin Higgs Models:} 
For Composite Twin Higgs models, radiative contributions from the top sector can be significantly smaller, and so the tadpole-induced EWSB structure allows the tuning to be substantially improved compared to the `irreducible' $\frac{f_H^2}{2v_H^2}$ tuning of a purely radiative potential. The result can be, for example, an untuned model with a global symmetry-breaking scale $f_H \sim1\TeV$ and colored top partners at 2 TeV.

In \Sref{sec:structure}, we discuss in more detail the structure of these models in the presence of a non-dynamical tadpole term. In \Sref{sec:radiativetuning}, we give concrete examples of top sectors and discuss the advantages of the additional tadpole contribution to the potential, including when the tuning can be substantially improved by the induced EWSB structure.  In \Sref{sec:SigmaDynamics}, we discuss the dynamics of the $\Sigma$ sector, demonstrating that a realistic strongly-coupled auxiliary sector preserves the improved tuning and that the dynamical scale of the $\Sigma$ sector may even arise from the Higgs sector itself. In \Sref{sec:expt} we discuss phenomenological constraints on the Higgs properties, extended Higgs or $\Sigma$ sector states, and top partners. Finally, we conclude in \Sref{sec:conclusion}.

\section{pNGB Models with Tadpoles}
\label{sec:structure}

In Composite Higgs models, the SM Higgs is identified with a pNGB in the coset $G/H$ of the spontaneously-broken global symmetry $G \rightarrow H$. We discuss how the presence of a tadpole term modifies the structure of the pNGB potential, naturally allowing $v_H \ll f_H$ with radiative contributions setting the scale for $m_h =125\GeV$.

\subsection{Minimal coset pNGB Higgs models} \label{sec:introA}

For the models relevant to our discussion, the radiatively-generated Higgs potential can be parameterized as~\cite{Bellazzini:2014yua}
\be
\label{eq:pngbpotential}
V(h) = \alpha f_H^4 \sin^2\left(\frac{h}{f_H}\right) + \beta f_H^4 \sin^4\left(\frac{h}{f_H}\right)
\ee
where $f_H$ is the scale of spontaneous $G$ breaking. For ${\alpha<0}$, EWSB with scale $v_H$ in the Higgs sector is triggered. The hierarchy compared to the global symmetry breaking scale is
\be
\label{eq:vev}
\frac{v_H^2}{f_H^2} = \sin^2\left(\frac{\langle h \rangle}{f_H}\right) = -\frac{\alpha}{2 \beta},
\ee
while the physical mass-squared is
\be
m_h^2 = 2 \beta f_H^2 \sin^2\left(\frac{2 \langle h \rangle}{f_H}\right) = -4\alpha f_H^2 \cos^2\left(\frac{\langle h \rangle}{f_H}\right)
\ee
and $m_h=125\GeV$ is realized for $\beta = \betaexp \simeq 1/32$.

A key point is that radiative contributions to the potential from the explicit $G$-breaking couplings of the SM generically generate $|\alpha| \gsim \beta$.  To realize a hierarchy $v_H \ll f_H$ requires $|\alpha| \ll \beta$, and in the case of a purely radiative potential this can only be arranged with a tuned cancellation among the different contributions to $\alpha$. Explicitly, assuming that the physics responsible for generating the required $\beta = \betaexp $ also generates a comparable contribution to $\alpha$, and taking $\frac{v_H^2}{f_H^2} \ll 1$, implies a tuning
\be
\label{eq:vftuning}
\frac{\delta \alpha}{\alpha} \gsim \frac{\beta}{2 \beta (v_H^2/f_H^2)} = \frac{f_H^2}{2 v_H^2}.
\ee

However, this tuning is not `irreducible'---it can be avoided by including an additional tadpole-like contribution to the potential.
The structure of the low-energy theory is that of `induced' EWSB \cite{Galloway:2013dma,Chang:2014ida}. In induced EWSB, the Higgs vev arises as a result of a coupling linear in the Higgs to another sector, $\Sigma$, that breaks the electroweak symmetry at $f_\Sigma \ll v_H$,
\be
V(H) \supset -\kappa^2H\cdot\Sigma + \hc
\ee
with $\vev{\abs{\Sigma}} = \frac{f_\Sigma}{\sqrt{2}}$.
If this additional sector were not present or did not acquire a vev, Higgs EWSB would not occur.  In the limit that the extra modes of the additional sector are decoupled, the dominant component of EWSB can be viewed as arising from an effective tadpole for the Higgs; we first focus on this case before returning to the dynamics of the $\Sigma$ sector in \Sref{sec:SigmaDynamics}.

For a composite Higgs model, we can parameterize the tadpole by a term $\gamma=\kappa^2 f_\Sigma / f_H^3$ in the non-linear realization,
\be
\label{eq:simpletadpoleV}
V(h) = - \gamma f_H^4 \sin\left(\frac{h}{f_H}\right) + \alpha f_H^4 \sin^2\left(\frac{h}{f_H}\right) + ...
\ee
such that
\be
\label{eq:tadpolemin}
\frac{v_H}{f_H} = \sin\left(\frac{\langle h \rangle}{f_H}\right) = \frac{\gamma}{2 \alpha}, \quad m_h^2 = 2 \alpha (f_H^2-v_H^2).
\ee
This mechanism requires $\alpha > 0$, such that $v_H = 0$ for $\gamma = 0$.
The tadpole perturbs the vacuum from $v_H=0$ and a small value of $\gamma$ naturally leads to $v_H \ll f_H$.
As such, the correct Higgs mass and vev can be achieved even with $\beta\ll\betaexp$.
Moreover, since $\gamma$ explicitly breaks $SU(2)_L$, a hierarchy $\gamma\ll\alpha$ is naturally preserved by radiative corrections. As long as radiative contributions to the mass-squared can be made naturally small, $\delta \alpha f_H^2 \lsim m_h^2$, the overall naturalness of the model can be improved.

\subsection{Twin Higgs Models}
\label{sec:twinduced}

Twin Higgs models extend the coset and low-energy content of the theory to preserve a spontaneously broken ${\mathbb Z}_2$ mirror symmetry by introducing new mirror top and gauge partners at the scale $v_B \sim f_H$. The restored ${\mathbb Z}_2$ symmetry is sufficient to cut off the quadratic sensitivity of the pNGB potential at the scale $f_H$ instead of the scale of the colored top partners, which are somewhat heavier.

The original twin Higgs model \cite{Chacko:2005pe,Chacko:2005un} consisted of an $SU(4)$-invariant potential
\be
\label{eq:twinHpot1}
V = - M^2 \left(\abs{H_A}^2 + \abs{H_B}^2\right) + \lambda \left(\abs{H_A}^2 + \abs{H_B}^2\right)^2,
\ee
where $H_{A, B}$ are doublets of weakly-gauged $SU(2)_{A, B} \subset SU(4)$, with a small $SU(4)$-violating but $\mathbb{Z}_2$-preserving quartic
\be
\label{eq:twinHpot2}
V \supset \delta \left(\abs{H_A}^4 + \abs{H_B}^4\right).
\ee
The parity exchanges $A$ and $B$. 
In strongly-coupled realizations a larger $SO(8)$ symmetry should be considered~\cite{Chacko:2005un,Geller:2014kta,Barbieri:2015lqa,Low:2015nqa}.\footnote{For a review of some of the flavor phenomenology of Composite Twin Higgs, see~\cite{Csaki:2015gfd}.}
When the approximate $SU(4)$ is spontaneously broken by a large vev $f_H \gg v_H$, there is an uneaten pNGB that is associated with the Higgs, which develops a potential proportional to explicit $SU(4)$ breaking.  Parameterizing
\be
\abs{H_A}^2 = \frac{f_H^2}{2} \sin^2\left(\frac{h}{f_H}\right), \quad \abs{H_B}^2 = \frac{f_H^2}{2} \cos^2\left(\frac{h}{f_H}\right),
\ee
one finds a potential for the light Higgs mode of the form of \Eref{eq:pngbpotential} with $\beta = -\alpha = \frac{\delta}{2}$.  The $\mathbb{Z}_2$ symmetry ensures that quadratically-divergent radiative contributions take the form $\Lambda^2 \left(\abs{H_A}^2 + \abs{H_B}^2\right)$, which is independent of the light Higgs field.

For $\delta > 0$, as for the IR contribution of a $\mathbb{Z}_2$-preserving top sector, the unbroken parity would enforce $\abs{H_A}^2 = \abs{H_B}^2 = \frac{f_H^2}{4}$.  In this case, achieving $\abs{H_A}^2 = \frac{v_H^2}{2} \ll \frac{f_H^2}{2}$ (associating the SM weak gauge group with $SU(2)_A$) requires explicit $\mathbb{Z}_2$ breaking.  In the original model, this was accomplished by a soft $\mathbb{Z}_2$-breaking mass term
\be
\Delta V = \Delta m^2 \left(\abs{H_A}^2 - \abs{H_B}^2\right),
\ee
giving an additional contribution to $\alpha \sim \frac{\Delta m^2}{f_H^2}$.  This contribution can be fine-tuned against the above contribution $\delta \alpha = -\frac{\delta}{2}$ to get the correct vev, but also results in the tuning described for the conventional models.  If the top sector generates the observed value of $\delta = 2 \betaexp$, these models exhibit a minimal tuning $\Delta \geq \frac{f_H^2}{2 v_H^2}$.

Introducing an EWSB-inducing sector can readily remove this tuning in the Twin case. We assume that, prior to EWSB, the $SU(4)$-breaking vev is stabilized at $\abs{H_A}^2 = 0, \abs{H_B}^2 = \frac{f_H^2}{2}$.  This can be achieved in the limit of unbroken $\mathbb{Z}_2$, for instance if $\delta < 0$, or due to the presence of a large $\Delta m^2 > \delta f_H^2$---we will return to the possible origin of the various terms in \Sref{subsec:uvcompletion}.  In addition, we include $\mathbb{Z}_2$-symmetric tadpole terms
\be
\Delta V = -\kappa^2  \left(\Sigma_A \cdot H_A + \Sigma_B \cdot H_B\right) + \hc
\ee
Below the scale $f_H$, the Higgs potential takes the form
\begin{align}
V(h) & = - \kappa^2 f_\Sigma f_H \left(\sin\left(\frac{h}{f_H}\right) + \cos\left(\frac{h}{f_H}\right)\right) \nonumber\\ &+ \alpha f_H^4 \sin^2\left(\frac{h}{f_H}\right) - \beta f_H^4 \sin^4\left(\frac{h}{f_H}\right)
\label{eq:tadpolePotentialTH}
\end{align}
such that the Higgs vev is determined by
\be
- 2 \beta \sin^3\left(\frac{\vev{h}}{f_H}\right) + \alpha \sin\left(\frac{\vev{h}}{f_H}\right) + \frac{\kappa^2 f_\Sigma}{f_H^3} \tan\left(\frac{\vev{h}}{f_H}\right) = \frac{\kappa^2 f_\Sigma}{f_H^3}.
\ee
For $\beta \sim \alpha$ and $f_H \gg v_H$, $\tan\left(\frac{v_H}{f_H}\right) \simeq \sin\left(\frac{v_H}{f_H}\right)$ and the cubic term can be approximately neglected.  So, the correct vev is simply achieved by the tadpole
\be
\kappa^2 f_\Sigma \simeq \frac{\alpha f_H^3 \sin\left(\frac{\vev{h}}{f_H}\right)}{1 - \sin\left(\frac{\vev{h}}{f_H}\right)} = \alpha f_H^2 v_H \left(1+\mathcal{O}\left(\frac{\vev{h}}{f_H}\right)\right).
\label{eq:tadpolevev}
\ee
Just as before, the tadpole allows the vev to be continuously perturbed away from the $v_H=0$ vacuum, giving a hierarchy $v_H \ll f_H$ without any tuning. 

It is interesting to note that, while the enlarged structure of the Twin Higgs due to the ${\mathbb Z}_2$ symmetry permits multiple possibilities for the unperturbed vacuum (\ie, with $\kappa^2 = 0$), this reduced tuning is unique to the model perturbing around $(\abs{H_A},\abs{H_B})=\left(0, \frac{f_H}{\sqrt{2}}\right)$ with ${\mathbb Z}_2$-symmetric tadpoles.
In principal, spontaneous or explicit ${\mathbb Z}_2$ breaking in the $\Sigma$ sector could give a tadpole only in the $A$-sector, $(f_{\Sigma_A},f_{\Sigma_B})=\left(\frac{f_\Sigma}{\sqrt{2}}, 0\right)$. But, in this case, the vacuum with $v_H \ll f_H$ reached by perturbing about $(\abs{H_A},\abs{H_B})=\left(0, \frac{f_H}{\sqrt{2}}\right)$ is always unstable to a global vacuum at $v_H = f_H$ reached from the unperturbed $(\abs{H_A},\abs{H_B})=\left(\frac{f_H}{\sqrt{2}},0\right)$ vacuum.  Moreoever, this is true even in the presence of higher-order terms that may induce a misalignment in the $SU(2)$ orientation of the $H_{A,B}$ and $\Sigma_{A,B}$ vevs.
Alternatively, \Rref{Beauchesne:2015lva} considered a similar model with the unperturbed vacuum instead at $(\abs{H_A},\abs{H_B})=\left(\frac{f_H}{2}, \frac{f_H}{2}\right)$ and a spontaneous ${\mathbb Z}_2$ breaking in the tadpole sector, $(f_{\Sigma_A},f_{\Sigma_B})=(0,f_\Sigma)$. The tadpole helps favor $\vev{\abs{H_A}} < \vev{\abs{H_B}}$, but obtaining a hierarchy $v_H \ll f_H$ is still a large perturbation away from the unperturbed vacuum, requiring a tuning of the tadpole against the parameters of the Higgs sector.

Thus $v_H\ll f_H$ can be obtained naturally in the Twin Higgs model with a tadpole $\gamma=\kappa^2 f_\Sigma/f_H^3$ that is protected against the radiative contributions generating the mass term $\alpha$.  Provided radiative contributions to $\alpha$ are sufficiently small, $\delta \alpha \sim \frac{m_h^2}{f_H^2}$, such a model can be considerably less tuned than the original Twin Higgs model.
We discuss the improved naturalness of induced EWSB for concrete examples of Twin and minimal Composite Higgs models in the next section.

\section{Radiative Tuning from the Top Sector}
\label{sec:radiativetuning}

The analysis of the preceding section establishes that a tadpole can decouple the hierarchy $v_H \ll f_H$ from the parameters in the Higgs sector, and hence from the scale of radiative corrections to those parameters.
At very least, this allows the correct Higgs vev and mass to be achieved without the requirement of generating a sufficiently large $\beta \gg \abs{\alpha}$. Moreover, as radiative contributions tend to produce $\abs{\alpha} \gsim \beta$, leading to the `irreducible' tuning $\Delta \gsim \frac{f_H^2}{2 v_H^2}$, induced EWSB permits the possibility of a more natural model than the minimal radiative models.
However, for tuning to be significantly improved, it is necessary that radiative contributions can be reduced relative to those of the minimal radiative models.
To identify cases in which this can occur, we will now discuss more concretely the form of the top sector and the size of radiative corrections for both minimal $SO(5)/SO(4)$ Composite (\Mfive) and Twin Higgs models.

An important constraint comes from realizing the top Yukawa coupling, which in Composite Higgs models gives a rough lower limit on the top partner masses compared to the global symmetry breaking scale, $m_\ast \gsim \frac{y_t f_H}{\sqrt{2}}$.  Note that, in a Twin Higgs model, this is the mass of the SM singlet $B$-sector top.  As a result, there is a direct lower bound on the tuning of $\alpha$ even after we have introduced the tadpole structure,
\be
\label{eq:radtoptuningmin}
\frac{\delta m_h^2}{m_h^2} \gsim \frac{3 y_t^4 f_H^2}{8 \pi^2 \left(\frac{v_H^2}{4}\right)} \simeq \frac{1}{6} \left(\frac{f_H^2}{2 v_H^2}\right).
\ee
By comparison, the `irreducible' tuning \Eref{eq:vftuning} resulted from the minimal size of radiative corrections necessary to obtain $\beta=\betaexp$ and a hierarchy $v_H \ll f_H$ in a purely radiative potential. So, the tadpole structure has the potential to substantially reduce this tuning, particularly if the size of the top sector radiative contributions can be reduced as much as \Eref{eq:radtoptuningmin} na\"ively suggests.

This possibility is especially pertinent in light of recent LHC results.
Current constraints on Higgs properties require $f_H \gsim 750 \GeV$.
Without extra tree-level contributions to the Higgs potential, this would imply that the minimal radiative models are tuned at the level of at least ${\cal O}(20\%)$, indicating a tension with the principle of naturalness. 
In contrast, constraints on colored top partners are $m_\ast \gsim 700 \GeV$, and are significantly weaker for the uncolored partners in TH.
For such values of $f_H$ and $m_\ast$, the radiative tuning can be still ${\cal O}(1)$, corresponding to an essentially untuned model.

Motivated by bounds on Higgs properties, we will fix $f_H=1\TeV$ ($\frac{f_H^2}{v_H^2} \simeq 16$) as a benchmark in this section with $f_\Sigma=70\GeV$ (giving $v_H=236\GeV$) for the MCHM and $f_{\Sigma_A} = 60\GeV$ for the Twin Higgs. As discussed in greater detail in \Sref{sec:expt}, this benchmark is at the edge of current limits.
For different values of $f_H$, the top partner masses can be scaled as $f_H$ and the associated tuning as $f_H^2$; the tuning from the top sector is insensitive to $f_\Sigma$ when $f_\Sigma \ll v_H$.
We require (to leading order in $\frac{v_H^2}{f_H^2}$), 
\be
\alpha_{\rm obs}= \alpha_0 + \delta\alpha \simeq \frac{(125\GeV)^2}{2 f_H^2} \simeq \frac{1}{8} \frac{v_H^2}{f_H^2}
\ee
to realize EWSB with the observed Higgs mass.
We can therefore estimate the tuning of the tadpole model
\be
\Delta = \frac{\partial\ln \alpha}{\partial \ln \alpha_0} = 1 - \frac{\delta \alpha}{\alpha_{\rm obs}}
\label{eq:tuningformula}
\ee
The radiative contribution from the top sector is often negative in the concrete models of the top sector we study. Sources of $\alpha_0$ from outside the top sector that can be used to tune against this negative contribution and achieve the $\alpha = \alpha_{\rm obs} > 0$ required for induced EWSB are discussed in \Sref{subsec:uvcompletion}.

In practice, we find that the tadpole mechanism in the $SO(5)/SO(4)$ model allows $m_h=125\GeV$ to be obtained with the minimal representations of the fermion partners (\Mfive) and a tuning of $\sim 10\%$. This is a significant improvement over the $\sim 1\%$ tuning exhibited by an \Mfive~model in which  $\beta=\betaexp$ is radiatively generated.
However, essentially because the top partners cannot be made lighter than $\sim 2 f_H$, the tuning is still dominated by the radiative top sector tuning even with an additional tadpole contribution. As such, the overall naturalness of the model is not necessarily improved compared to the $\frac{f_H^2}{2v_H^2}$ tuning of an extended radiative potential.
By comparison, in Twin Higgs models, the neutral top partners can be sufficiently light to realize the lower limit of \Eref{eq:radtoptuningmin}, such that the induced structure offers a substantial improvement in naturalness over any radiative model.

\subsection{$SO(5)/SO(4)$ with a Minimal Top Sector}
\label{subsec:5plus1}

In the partial compositeness framework, the mixings of the elementary and composite fermions generate the top Yukawa coupling. The embedding of the top partners in the global symmetry group determines the form of the radiative correction, and the minimal \Mfive~has composite fermionic partners (including colored vector-like top partners) in the $5=4+1$, $\psi=(\psi_{4}^i, \psi_{1})$ and $\psi^c=({\psi^c_{4}}^i, {\psi^c_{1}})$~\cite{Agashe:2004rs,Bellazzini:2014yua,Panico:2015jxa}, with $q_L$ mixing as a $5 =4+1$ and $t_R$ as a singlet. For this embedding, only the mixings of $q_L$ explicitly break the global symmetry. Contributions to $\alpha$ are quadratically sensitive to the top partner mass, while contributions to $\beta$ are only logarithmically sensitive, yielding $\abs{\alpha} \sim \frac{m_\ast^2}{f_H^2} \abs{\beta}$.

This is an interesting case to apply the tadpole mechanism of EWSB as $\abs{\alpha} > \abs{\beta}$ implies that, if the top sector is responsible for radiatively generating the observed value of $\beta=\betaexp$, the tuning is considerably worse than the minimal tuning, $\Delta \gg \frac{f_H^2}{2 v_H^2}$.
Induced EWSB removes the restriction of obtaining $\beta=\betaexp$ from the top sector. The top partner mass scale $m_\ast$ is then only restricted by direct experimental limits and the requirement of realizing the top Yukawa coupling, and we can directly study how the tuning depends on $m_\ast$.

In a two-site model~\cite{Contino:2006nn,Panico:2011pw} for this composite sector, the radiative contributions to the Higgs potential can be calculated directly and parameterized in terms of two top partner mass scales, $m_1$ and $m_4$, and the mixing angles $\sin\theta_{L,R}$ of the top quark with the composites $\psi_A,\psi_A^c$. To leading order in $\frac{v_H}{f_H}$, the $SU(2)_L$-doublet top partners have masses $m_4$ and $M_4=m_4 / \cos\theta_L$, and the $SU(2)_L$-singlet top partner has mass $M_1=m_1/\cos\theta_R$. The Yukawa coupling is
\be
\label{eq:Yukawa51}
y_t = \frac{m_4}{f_H}\sin\theta_L \sin\theta_R
\ee
to leading order, which requires $m_4 \gsim f_H$, and gives a lower bound ${\frac{M_4}{f_H} \gsim \frac{2}{\sin\theta_R}}$ for the top partner mixing with the elementary $t_L$. For numerical results, we use $y_t=y_{t,SM} (v/v_H)$, where $y_{t,SM}$ is the $\overline{MS}$ value at $1\TeV$.
The full definition of the two-site model and the radiative Higgs potential is given in \Aref{app:radiative}.  In the limit of a fully composite $t_R$, $\sin\theta_R=1$ and
\be
\delta \alpha = -\frac{3 y_t^2}{16\pi^2}\frac{M_4^2}{f_H^2}\left(1 + \log\left(\frac{\mu^2}{M_4^2}\right)\right)
\ee
The one-loop quadratic divergences are cut-off, but a residual logarithmic scale-dependence remains associated with the scale $\mu$ of the next set of top partner resonances \cite{Panico:2015jxa}. For concreteness, we set $\mu=3 M_4$.

\begin{figure}
\centering
\includegraphics[width=0.499\textwidth]{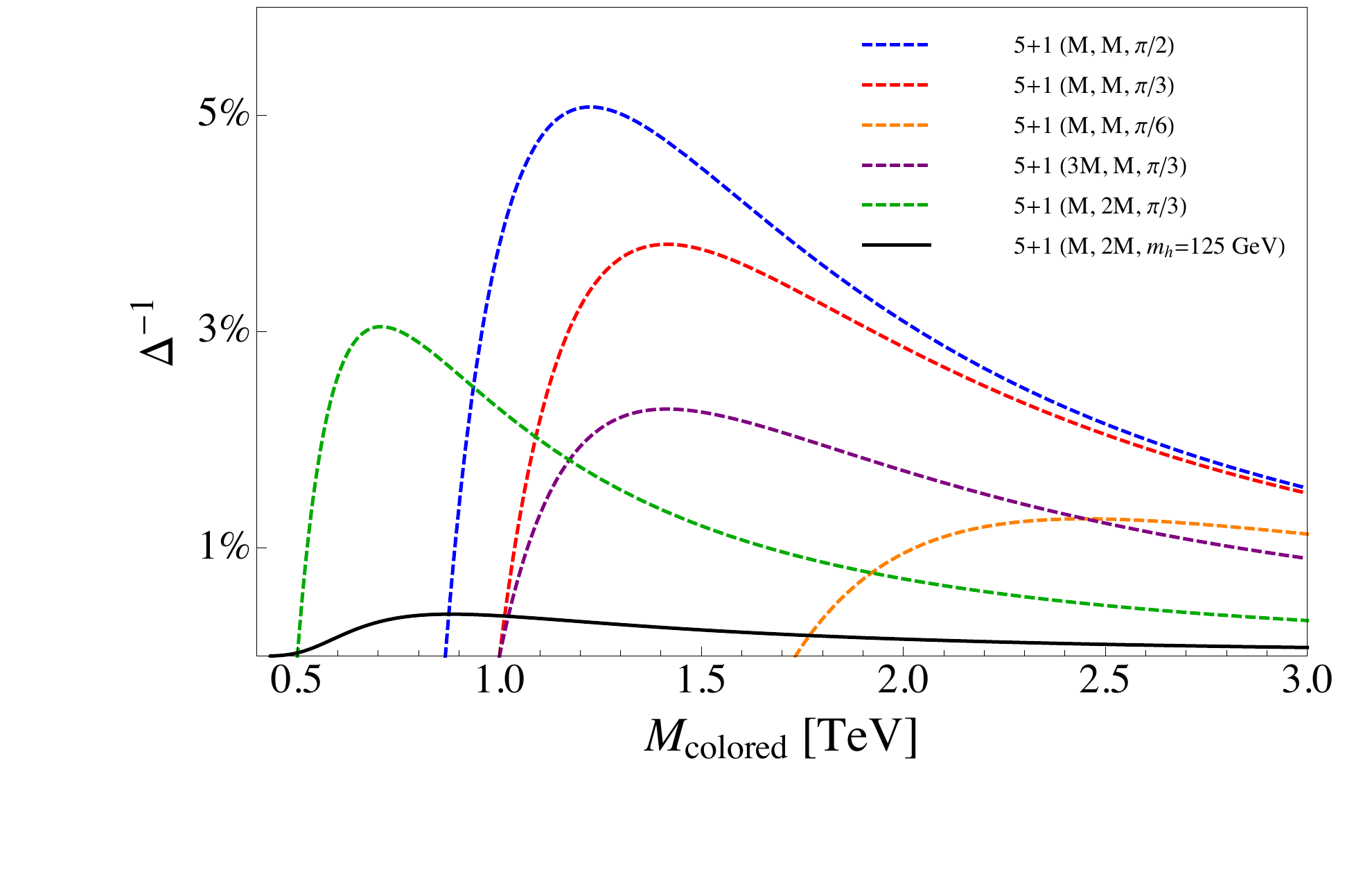}
\caption{
\label{fig:tadpoletuningSO5}
Top sector radiative tuning in the $SO(5)/SO(4)$ 5+1 model (or \Mfive) with a tadpole as a function of the lightest colored top partner mass $M_{\rm colored}$ for $f_H = 1\TeV$. Dashed curves correspond to different choices of $(M_1, m_4, \theta_R)$, as listed in the legend. For comparison, the black solid line corresponds to \Mfive~without a tadpole (\ie, with $\beta=\betaexp$ generated by large $q_L$ compositeness, determining $\theta_R$).
}
\end{figure}

\Fref{fig:tadpoletuningSO5} shows the tuning in this tadpole model as a function of the lightest top partner mass for several different sets of parameters $(M_1, m_4, \theta_R)$, with $\sin\theta_L$ determined from~\Eref{eq:Yukawa51}.
For comparison, we also show the tuning for the \Mfive~model without a tadpole in which the minimal top sector generates $\beta=\betaexp$ radiatively to give $m_h = 125 \GeV$. Achieving sufficiently large $\beta = \betaexp$ requires an increase in $q_L$ compositeness, such that the Higgs experiences more explicit breaking from $y_L > y_t$.\footnote{Similar to raising $m_h$ via large $A$-terms in the MSSM---the increased explicit symmetry breaking enhances the quartic, but also results in more tuning.} This in turn leads to more tuning.
A model exhibiting top partners with masses $\gsim f_H$ and a tadpole contribution to the potential can be significant more natural (with tuning reduced by ${\cal O}(5-10)$) than the \Mfive~with $\betaexp$ generated by the minimal top sector.

Because the top partners cutting off the quadratic sensitivity are always heavier than $\sim 2f_H$, the radiative tuning from the top sector in this model is always worse than $\frac{f_H^2}{2 v_H^2}$. 
As such, one can also consider alternatives to induced EWSB. For example one could include additional non-minimal radiative contributions giving $\beta = \betaexp$ with $\abs{\alpha} \sim \beta$, which would not substantially increasing the tuning. For example, the `maximally natural' top sector of the \Mfive~model can be supplemented by additional radiative contributions to the potential with $\abs{\delta \alpha} \simeq \abs{\delta \beta}$ from large $\tau_R$ compositeness.\footnote{This can be accomplished, \eg, in the framework of~\cite{Carmona:2014iwa}.} 
Another possibility is an extended top sector, for example the ${\rm MCHM}_{14+1}$ model gives $\abs{\alpha} \sim \abs{\beta}$ and may be able to radiatively realize $\beta=\betaexp$ in the region of parameter space with $m_\ast \sim f_H$. Therefore, in $SO(5)/SO(4)$ models, other equally natural realizations may exist. But, the tadpole mechanism is attractive for preserving the minimal partial compositeness partner realization.

In the following section, however, we study a Twin Higgs model where the quadratic sensitivity is cut off below the scale $f_H$, and the tadpole model can thus substantially improve the tuning compared to the $\frac{f_H^2}{2 v_H^2}$ tuning obtainable in a purely radiative model.

\subsection{Twin Higgs}
\label{subsec:twintoptuning}

The quadratic sensitivity of $\alpha$ to the top sector in Twin Higgs models is cut off by the twin top at $m_{t_B}\simeq \frac{y_t f_H}{\sqrt{2}}$, but a logarithmic sensitivity remains to the scale~$M_T$ of new \emph{colored} top partners that restore the full global symmetry in the top sector, 
\be 
\delta \alpha \simeq - \frac{3 y_t^4}{32\pi^2}\log \frac{M_T^2}{m_{t_B}^2}.\label{eq:LeadingLogAlpha}
\ee
We will study two concrete models of Twin top sectors to determine the degree to which light colored top partners can lower the radiative tuning of the tadpole potential with respect to the minimal $\frac{f_H^2}{2 v_H^2}$ tuning of the purely radiative potential.
Current direct experimental bounds require only $M_T \gsim 700~\GeV$ and will not significantly constrain the naturalness of these models. 
However, realizing the observed top Yukawa coupling and including threshold contributions to \Eref{eq:LeadingLogAlpha} again gives a lower bound on the tuning.

\begin{figure}
\centering
\includegraphics[width=0.47\textwidth]{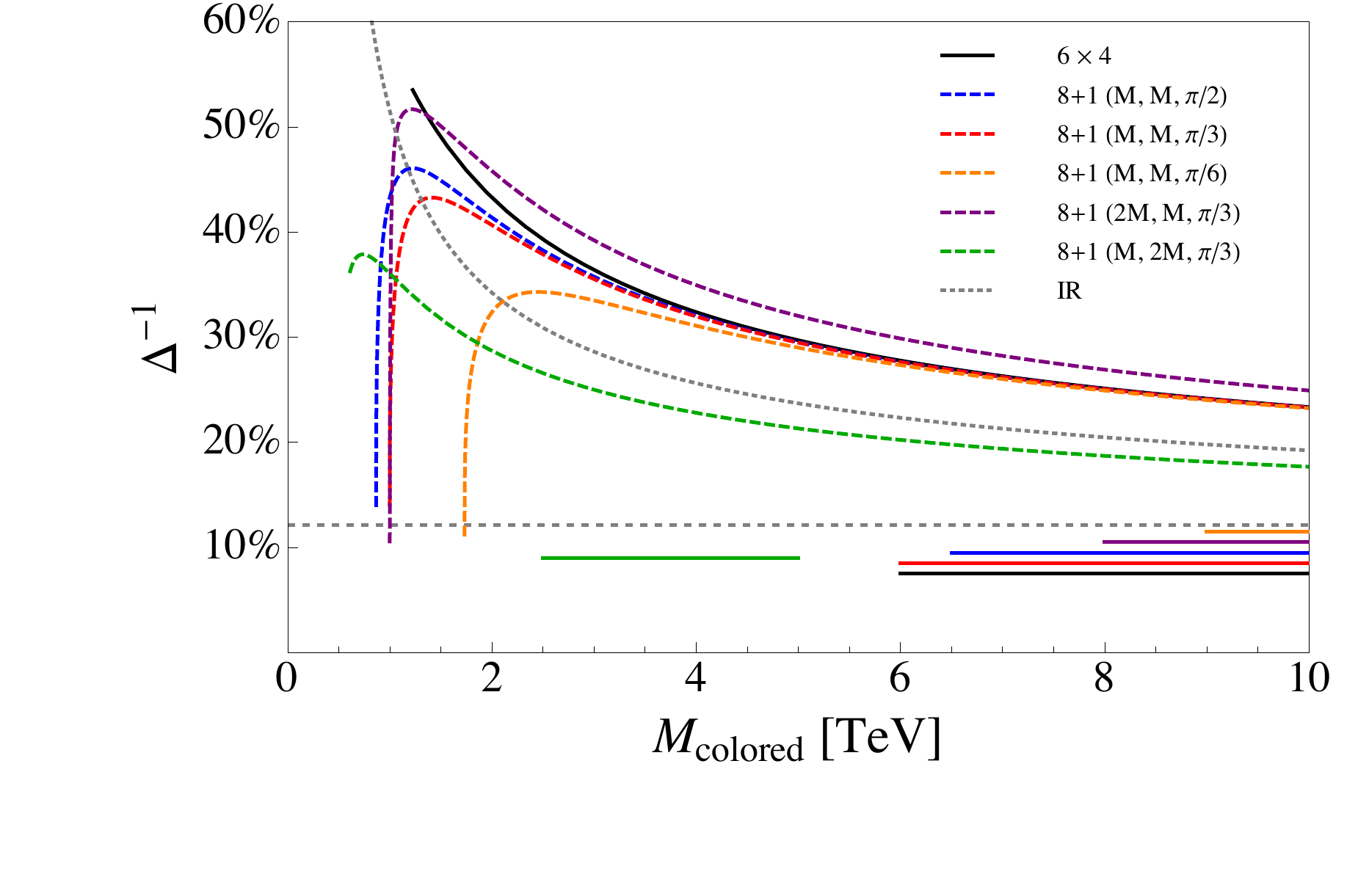}
\caption{
\label{fig:tadpoletuning}
Top sector radiative tuning in a Twin Higgs model with a tadpole as a function of the lightest colored top partner mass $M_{\rm colored}$ for $f_H=1\TeV$. Dotted gray is the estimated tuning from the pure $t_B$ contribution of \Eref{eq:LeadingLogAlpha}. Solid black is the $6\times4$ model, while dashed, colored curves correspond to the $8+1$ model with $(M_1, m_7, \theta_R)$ as listed. For comparison, the horizontal dotted gray line corresponds to the minimal tuning $\frac{2 v_H^2}{f_H^2} \simeq 10\%$ of the radiative quartic potential, with horizontal lines indicate the top partner mass range which can radiatively generate $\beta=\betaexp$ (saturating this tuning) within theoretical uncertainty.}
\end{figure}

The results are summarized in \Fref{fig:tadpoletuning}, which compares the tuning in several models to the logarithmic estimate \Eref{eq:LeadingLogAlpha}.
Unsurprisingly, we find that the minimal tuning occurs for top partners with masses roughly just above the smallest possible value required to realize the top Yukawa, $M_T \simeq \sqrt{2} f_H \sim m_{t_B}$. For these values, induced EWSB can reduce tuning by a factor of $\sim 5$ relative to the minimal $\frac{2 v_H^2}{f_H^2} \sim 10\%$ tuning of the radiative quartic potential.

\subsubsection*{$6\times4$ Top Sector}

\Rref{Chacko:2005pe} proposed completing the top sector by extending  $(Q_A, Q_B)$ into a $Q=(6,\overline{4})$ of $SU(6) \times SU(4)$, with $SU(3)_{c, A} \times SU(3)_{c, B} \subset SU(6)$. $Q$ contains new top partners $(\tilde{q}_A, \tilde{q}_B)$ required to restore the global symmetry in the $(3_A, 2_B)$ and $(3_B, 2_A)$ representations. The Yukawa coupling $y H Q U$ respects the $SU(4)$ symmetry and the exotic mixed states can be lifted by soft $SU(4)$-breaking vector-like masses $M(\tilde{q}_A \tilde{q}^c_A + \tilde{q}_B \tilde{q}^c_B)$. We will refer to this as the `$6\times4$' model.

To leading order in $\frac{v_H}{f_H}$, the colored top partner mass is $M^2_{T_A} = M^2 + \frac{y^2  f_H^2}{2}$ while the uncolored mirror top and top partner have masses $m^2_{t_B} = \frac{y^2 f_H^2}{2}$ and $M^2_{T_B} = M^2$ respectively.\footnote{Our normalization of $f_H$ differs by a factor of $\sqrt{2}$ from \Rref{Chacko:2005pe}} The coupling $y$ is related to the top Yukawa coupling as $y^2 = y_t^2 \left(1-\frac{y_t^2 f_H^2}{2 M^2}\right)^{-1}$, such that there is a minimal value for the colored top partner mass $M_{T_A} \geq \sqrt{2} y_t f_H$.
The radiative contribution to $\alpha$ is 
\be
\delta \alpha = \frac{3}{16\pi^2}\frac{y^2 M^2 / f_H^2}{M^2 - \frac{y^2 f_H^2}{2}}\left(M^2 \log \frac{M^2_{T_A}}{M^2_{T_B}} - \frac{y^2 f_H^2}{2} \log \frac{M^2_{T_A}}{m^2_{t_B}}\right). \label{eq:alphaClassic}
\ee
We evaluate \Eref{eq:alphaClassic} using the SM $\rm\overline{MS}$ value of the top mass at $\mu= m_{t_B} \simeq 700\GeV$. 

\Fref{fig:tadpoletuning} shows the radiative tuning due to this top sector. 
Also shown in \Fref{fig:tadpoletuning} is the approximate range of colored top partner mass $M \sim 10 \TeV$ that gives $\beta=\beta_{\rm SM}$ and would saturate the $2v_H^2/f_H^2$ tuning in the absence of the tadpole (we estimate the theoretical uncertainty by varying the top Yukawa coupling between its $\overline{\rm MS}$ values at $\mu=m_t$ and $\mu=m_{t_B}$).
For $M < 3f_H$, the tuning becomes considerably less than the `irreducible' tuning exhibited when $\beta=\betaexp$. At $M \simeq f_H$, the coupling $y$ becomes large and the tuning begins to worsen. For $M \gg f_H$, $\delta\alpha$ matches the expected logarithmic behavior \Eref{eq:LeadingLogAlpha}.
The minimally-tuned tadpole potential can permit significantly lower colored top partner masses, and correspondingly substantially reduced tuning.

\subsubsection*{$8+1$ Top Sector}

\Rrefs{Barbieri:2015lqa,Low:2015nqa} studied pNGB Twin Higgs models based on an $SO(8)/SO(7)$ coset with a partially composite top sector, similar to those studied in the MCHM~\cite{Agashe:2004rs,Bellazzini:2014yua,Panico:2015jxa} and above. In particular we focus on the model studied in \Rref{Low:2015nqa} with $q_L$ embedded in an $8 =7+1$, $t_R$ in a singlet, and composite top partners $\psi_A=(\psi_{7,A}^i, \psi_{1,A})$ and $\psi_B^i = (\psi_{7,B}^i, \psi_{1,B})$ in a $(3_A,8)$ and $(3_B,8)$ respectively. This is the Twin analog of the \Mfive~model. 

In a two-site model for this composite-sector, the radiative contributions to the Higgs potential can be calculated directly and parameterized in terms of two top partner mass scales $m_1$ and $m_7$ and the mixing angles $\sin\theta_{L,R}$ of the top quark with the composites $\psi_A,\psi_A^c$. To leading order in $\frac{v_H}{f_H}$, the colored $(3_A,2_A)$ top partners are at masses $m_7$ and $M_7=m_7 / \cos\theta_L$, and the $(3_A,2_B)$ top partners are at a mass $M_1=m_1/\cos\theta_R$. The Yukawa coupling is 
\be
\label{eq:Yukawa81}
y_t = \frac{m_7}{f_H}\sin\theta_L \sin\theta_R
\ee
to leading order, which requires $m_7 \gsim f_H$. 

The full definition of the two-site model and expressions for the radiative corrections are described in  \Aref{app:radiative} following \Rref{Low:2015nqa}. In the Twin model the contributions to $\alpha$ are only logarithmically sensitive to the colored top partner masses, and therefore the residual scale dependence found in the two-site 5+1 model is absent.

\Fref{fig:tadpoletuning} shows the tuning of the tadpole potential for the $8+1$ model in terms of the parameters $(M_1, m_7, \theta_R)$, with $\sin\theta_L$ fixed by the top Yukawa, \Eref{eq:Yukawa81}. 
Again, we highlight the top partner masses that would give $\beta=\betaexp$ and so saturate the $\frac{f_H^2}{2v_H^2}$ tuning (\ie, in the absence of the tadpole). We observe an improvement in tuning by a factor of $\sim 5$ is possible with the tadpole.
The improvement is substantial over most of the parameter space with $m_7 \lsim 3f_H$, but the tuning begins to worsen as the physical mass $M_7$ gets large at the lower range of $m_7$.
Note that the improvement in tuning by a factor of $\sim 10$ compared to the $5+1$ model can be understood as a result of uncolored top partners cutting off the quadratic sensitivity at a substantially lower scale than that at which colored top partners can appear. 

\section{Dynamical Auxiliary Sectors and UV Completions}
\label{sec:SigmaDynamics}

So far, we have considered a tadpole that arises due to an unspecified auxiliary sector exhibiting an $SU(2)_L$-breaking vev $f_\Sigma$. However, the dynamics of the auxiliary sector are also relevant. For instance, the auxiliary sector experiences back-reaction from the non-zero Higgs vev, and it is important to ensure that this does not destabilize the auxiliary sector or lead to hidden tuning.
Meanwhile, any explicit $G$-breaking present in the auxiliary sector may be communicated to the Higgs sector.

The presence of an additional sector containing an electroweak doublet also leads to modifications of Higgs properties and novel states that may be produced at colliders.
Notably, a second doublet gives rise to additional charged and pseudoscalar Higgses, $H^\pm$ and $A$ respectively, similar to those of a fermiophobic/type-I two Higgs doublet model (in which only a single doublet couples to fermions).
Thus, the largely SM-like nature of the Higgs and the non-observation of BSM states at the LHC constrains the dynamics of the auxiliary sector.
Overall, the auxiliary sector must exhibit certain properties in order to remain stable against back-reaction, to stay consistent with experimental measurements, and to preserve the improved naturalness of the model.

One important question is whether the auxiliary sector is weakly- or strongly-coupled---\ie, is $\Sigma$ elementary or composite?
Several pieces of evidence point to a strongly-coupled auxiliary sector (as in, \eg, Bosonic Technicolor).
First, experimental constraints on Higgs couplings require $f_\Sigma \ll v_H$.
So, for $v_H \ll f_H$, \Eref{eq:tadpolemin} implies
\be
\label{eq:kappaSize}
\frac{\kappa^2}{f_\Sigma^2} = \frac{m_h^2 v}{f_\Sigma^3} \approx 10 \left(\frac{70\GeV}{f_\Sigma}\right)^3.
\ee
This is very similar to the size $\kappa^2 \sim 4 \pi f_\Sigma^2$ suggested by na\"ive dimensional analysis for a strongly-coupled auxiliary sector with an $\mathcal{O}(1)$ weak coupling to the Higgs sector.
Second, large couplings help stabilize $f_\Sigma \ll v_H$ against large back-reaction when the Higgs field acquires its vev.
Finally, large couplings raise the mass scale of the resonances associated with the auxiliary sector, explaining their non-observation thus far at the LHC.

A second issue is that the $\Sigma$ sector need not respect the approximate global symmetry $G$---in fact, explicit $G$-breaking in the $\Sigma$ sector can avoid additional light modes and may reduce its susceptibility to back-reaction. The details of the $\Sigma$ sector determine how this explicit breaking is communicated to the Higgs sector. The low-energy form of the coupling $\kappa^2 H \cdot \Sigma$ is a soft breaking of $G$ in the Higgs sector, and so the contributions to the pNGB potential will be under control even for a strongly-coupled auxiliary sector that generates important higher-order terms. However, for some UV completions of the $\Sigma$ sector there can be further UV-sensitive contributions to the pNGB potential.

In this section, we shall explore the structure of the auxiliary sector, beginning first with a linear model.
As the above constraints likely imply strong coupling, this model is more useful for developing intuition (\ie, in the large self-coupling limit) than it is realistic.
We shall subsequently discuss strongly-coupled auxiliary sectors, focusing on the additional higher-order operators between the Higgs and $\Sigma$ sectors we expect in this scenario.
While these operators may have interesting implications, the qualitative features of the model remain unchanged.
Finally, we will highlight some additional UV considerations relevant for models that attempt to address the origin of the two sectors.

\subsection{Linearly-Realized Auxiliary Sectors}
\label{sec:linearsig}

An effective theory analysis has previously been carried out in the context of a simplified model of induced EWSB with a single Higgs doublet coupled to a linearly-realized $\Sigma$ doublet in \cite{Galloway:2013dma}.
They confirmed that it was possible to achieve a stable vacuum with $f_\Sigma < v_H$ and, as the tadpole limit is approached, tuning does indeed become small.
Here, we extend this analysis to the case of the MCHM and Twin Higgs scenarios.
While the requirement of strong coupling limits the validity of a linear description of the auxiliary sector, this approach allows us to investigate the back-reaction, tuning and impact of $\Sigma$-sector $G$-breaking described above, as well as the form of the tadpoles generated. We will discuss in more generality the strongly-coupled case in the following subsection.

\subsubsection{Composite Higgs}

Starting with the $SO(5)/SO(4)$ case, we take the $\Sigma$ sector to be a simple linear model,
\be 
\label{eq:lsmpotential}
V_\Sigma = -\Lambda_\Sigma^2 \abs{\Sigma}^2 + \delta_\Sigma \abs{\Sigma}^4,
\ee
which only realizes the custodial $SO(4)$ symmetry.
In the absence of a coupling to the Higgs, $SO(4)$ is spontaneously broken at scale $f_\Sigma^2 = \frac{\Lambda_\Sigma^2}{\delta_\Sigma}$.
The Higgs and auxiliary sectors are linked by a $B\mu$-type term,
\be
V \supset -\kappa^2 \Sigma^\dagger H + \hc,
\ee
producing the necessary EWSB tadpole.  In addition, this term explicitly breaks $SO(5)_H \times SO(4)_\Sigma \rightarrow SO(4)$, both giving mass to the extra Higgs states $m_A^2 \simeq m_{H^\pm}^2 \propto \kappa^2$ and inducing $SU(2)_L$-alignment between $\vev{H}$ and $\vev{\Sigma}$.

To estimate the impact of back-reaction on the auxiliary sector, we focus on the neutral CP-even states, expanding about the unperturbed $\Sigma$ vacuum $\abs{\Sigma} = \frac{f_\Sigma + \sigma}{\sqrt{2}}$ and treating the Higgs pNGB as a background field.  This gives a quadratic potential
\be
V_\Sigma = \Lambda_\Sigma^2 \sigma^2 - \kappa^2 (f_H s_h) \sigma.
\ee
The effective tadpole for $\sigma$ shifts the $\Sigma$-sector EWSB vev
\be
\label{eq:sigmavevshift}
\vev{\sigma} \simeq \frac{\kappa^2 f_H \sin\left(\frac{v_H}{f_H}\right)}{2 \Lambda_\Sigma^2 }.
\ee
The auxiliary sector minimization condition combined with \Eref{eq:kappaSize} implies
\be
\frac{\vev{\sigma}}{f_\Sigma} \simeq \frac{m_h^2 v_H^2}{\delta_\Sigma f_\Sigma^4} \simeq 0.5 \left(\frac{4 \pi^2}{\delta_\Sigma}\right) \left(\frac{70 \GeV}{f_\Sigma}\right)^4.
\ee
So, the EWSB vev in the $\Sigma$ sector does receive a correction due to back-reaction from the Higgs vev, but this effect is suppressed in the strong coupling regime when $\delta_\Sigma$ is large.\footnote{For our chosen normalization of the quartic, nonperturbative self-coupling corresponds to $\delta_\Sigma \rightarrow 4 \pi^2$.} In particular, that the shift in $\vev{\Sigma}$ is relatively small in this regime indicates that back-reaction does not result in additional tuning.

Meanwhile, the Higgs experiences explicit $SO(5)$-breaking in addition to the tadpole through its interactions with $\sigma$. In this simplified picture, this breaking can be viewed as communicated via mixing of the CP-even states, which induces higher-order operators in the pNGB potential.  It is useful to define $\epsilon = \frac{\kappa^2}{2\Lambda_\Sigma^2}$ to parameterize the mixing angle of the Higgs pNGB and $\sigma$,
\be
\epsilon \simeq 0.14 \left(\frac{4 \pi^2}{\delta_\Sigma}\right) \left(\frac{70 \GeV}{f_\Sigma}\right)^4.
\ee
Again, these effects are suppressed in the large-coupling limit.
Integrating out $\sigma$ gives rise to new terms in the pNGB potential, including
\be
\label{eq:VhSigmin}
V_h \supset - \epsilon^2 \Lambda_\Sigma^2 f_H^2 s_h^2 \simeq - \frac{\kappa^4}{4 \Lambda_\Sigma^2 f_H^2} f_H^4 s_h^2
\ee
corresponding to a contribution to $\alpha$
\be
\abs{\frac{(\delta \alpha)_\Sigma}{\alpha}} \simeq 0.5 \left(\frac{4 \pi^2}{\delta_\Sigma}\right) \left(\frac{70 \GeV}{f_\Sigma}\right)^4.
\ee
In the strong-coupling limit, this effect is of similar size to the experimentally-required value of $\alpha$, and therefore does not induce additional tuning.  Higher-order terms are suppressed by powers of mixing between the Higgs and $\Sigma$ sector, but can be relevant for the phenomenology of the extra Higgs states, as will be discussed in \Sref{sec:strongcoupling}. 

This analysis indicates that the dynamics of the auxiliary sector do not disrupt the leading-order description of a Higgs pNGB with positive mass term ($\alpha > 0$) and EWSB induced by a tadpole as in \Sref{sec:structure}, particularly in the strong-coupling limit required by experimental constraints.  Back-reaction and explicit $SO(5)$-breaking lead to at most ${\cal O}(1)$ shifts to $(f_\Sigma, \alpha)$, and so for strong-coupling induce no additional tuning in either sector.

\subsubsection{Twin Higgs}

The Twin Higgs case is similar to the $SO(5)/SO(4)$ scenario described above except with one important difference, namely that the ${\mathbb Z}_2$ requires that both of the scales in the $H$ sector, including $v_B \sim f_H \gg v_H$, couple to the auxiliary sector. This causes a larger perturbation in the $\Sigma$ sector, although such perturbations can still be sufficiently small to avoid tuning or destabilization of the auxiliary sector. Moreover, the additional interactions between sectors may offer some intriguing opportunities, including generation of the required $\alpha > 0$, dynamical generation of the hierarchy $f_\Sigma \ll v_H \ll f_H$ and complete $SU(2)_B \times U(1)_B$-breaking. 

Extending the potential \Eref{eq:lsmpotential} to the Twin Higgs case, we consider a ``Twin Sister'' model\footnote{A twinning of the Sister Higgs~\cite{Alves:2012ez}.} with
\begin{align}
V_\Sigma & \supset - \Lambda_\Sigma^2 \left(\abs{\Sigma_A}^2 + \abs{\Sigma_B}^2\right) + \lambda_\Sigma \left(\abs{\Sigma_A}^2 + \abs{\Sigma_B}^2\right)^2 \nonumber \\
& \qquad + \delta_\Sigma \left(\abs{\Sigma_A}^4 + \abs{\Sigma_B}^4\right).
\end{align}
The Higgs sector is of the same form as given in \Erefs{eq:twinHpot1}{eq:twinHpot2} with $\delta \ll \lambda$ giving the approximate $SU(4)_H$ symmetry.
For simplicity, we take $\delta_\Sigma \gg \lambda_\Sigma$ and treat the $\lambda_\Sigma$ term coupling the $\Sigma_A$ and $\Sigma_B$ sectors as a perturbation. The unperturbed vev is then $f^2_{\Sigma_{A,B}} = \frac{\Lambda_\Sigma^2}{\delta_\Sigma}$, and $SU(4)_\Sigma$ is strongly broken to $SU(2)_{\Sigma_A} \times SU(2)_{\Sigma_B}$. 
The $H$ and $\Sigma$ sectors are again linked by a $B\mu$-type term, 
\be
V \supset -\kappa^2 \left(\Sigma_A^\dagger H_A + \Sigma_B^\dagger H_B + \text{h.c.}\right),
\ee
which is an explicit soft breaking of the $SU(2)_{\Sigma_A} \times SU(2)_{\Sigma_B} \times SU(4)_H$ global symmetry to the gauge and discrete symmetry $SU(2)_A \times SU(2)_B \times {\mathbb Z}_2$.

Following the same strategy of integrating out the $\Sigma$ sector, we have the leading quadratic terms
\begin{align*}
V_\Sigma & = \Lambda_\Sigma^2 (\sigma_A^2 + \sigma_B^2) - \kappa^2 f_H (s_h\sigma_A + c_h \sigma_B) \\ & \qquad + \lambda_\Sigma f_\Sigma^2 \sigma_A \sigma_B,
\end{align*}
where we have elided terms proportional to $\lambda_\Sigma$ that do not couple the $\Sigma_A$ and $\Sigma_B$ sectors.
The $B$-sector vev is shifted by
\begin{align}
\frac{\vev{\sigma_B}}{f_\Sigma} & = \epsilon \frac{f_H \cos\left(\frac{v}{f_H}\right)}{f_\Sigma} \nonumber \\
& \simeq 2 \left(\frac{4 \pi^2}{\delta_\Sigma}\right) \left(\frac{70 \GeV}{f_\Sigma}\right)^4 \left(\frac{1 \TeV}{f_H}\right)\,.
\end{align}
As anticipated, if there is a hierarchy $f_H\gg v_H$, this can be an $\mathcal{O}(1)$ perturbation even as $\delta_\Sigma$ approaches strong coupling. 

Likewise, there can be significant contributions to the Higgs potential.  The leading contribution present in the $SO(5)/SO(4)$ case, \Eref{eq:VhSigmin}, is cancelled because of the ${\mathbb Z}_2$ Twin protection. The ${\mathbb Z}_2$ breaking shift in $\vev{\sigma}$ captured by the $\sigma^3, \sigma^4$ terms, which give
\be
V_h \supset \delta_\Sigma f_\Sigma \epsilon^3 \left(s_h^3 + c_h^3\right) + \frac{\delta_\Sigma}{4} \epsilon^4 \left(s_h^4 + c_h^4\right)
\ee
potentially producing contributions $\abs{\frac{(\delta \alpha)_{\Sigma}}{\alpha}} \simeq {\cal O}({\rm few})$.
In particular, as $f_{\Sigma_B}$ is unconstrained by experiment, it can be somewhat larger than $f_{\Sigma_A}$, such that contributions to the Higgs mass can be somewhat enhanced.
This indicates the possibility that Higgs couplings to the auxiliary sector may be the source of the required $\alpha > 0$.

The back-reaction and mixing contributions remain comparable to the required $f_\Sigma$ and $\alpha$. So, while they make a complete analysis of the potential somewhat more complicated, they do not induce additional tuning.
We have confirmed this is the case with a full numerical study of the potential. Overall, our results are consistent with those of \cite{Galloway:2013dma}; it is possible to achieve ${\cal O}(1)$ tuning and a stable vacuum with $f_{\Sigma_A} \lsim f_{\Sigma_B} \ll v_H \ll f_H$.  A realistic auxiliary sector likely exhibits approximately $\mathbb{Z}_2$-symmetric vevs, a large explicit breaking of the global symmetry (\ie, $\delta_\Sigma \gsim \lambda_\Sigma$), and strong coupling.

So far, we have ignored the role of $\lambda_\Sigma$. Treating $\lambda_\Sigma$ as a perturbation, a shift in the $\Sigma_A$ vev is also generated at leading order as a result of $\vev{\sigma_B}$, $\vev{\sigma_A} = \frac{\lambda_\Sigma}{2\delta_\Sigma} \vev{\sigma_B}$. Clearly, for a more generic potential with $\lambda_\Sigma \sim \delta_\Sigma$, both $\Sigma_A$ and $\Sigma_B$ can experience large perturbations due to the $f_H$ tadpole.
This raises the interesting possibility that the hierarchy and coincidence of scales is generated by a ``waterfall'' of induced breakings originating from $f_H$.  For instance, in the limit that $\Lambda_\Sigma = 0$, the scale of the $\Sigma$ sector is set completely by the large $B$-sector tadpole from $f_H$.
While $\Lambda_\Sigma = 0$ is unnatural in the linear sigma model, this serves as a useful toy model for a strongly-coupled model where the scale of a conformal $\Sigma$ sector may be triggered by the coupling to the $H$ sector, as we will discuss shortly. Then $\delta_\Sigma^{3/2} f^3_{\Sigma_B} \simeq \kappa^2 f_H$, and a term $\lambda_\Sigma < 0$ can trigger the breaking in the $A$-sector. This waterfall of breaking then feeds back into the $H_A$ sector through the EWSB-inducing tadpole.

The linear sigma model nicely captures the back-reaction on the $\Sigma$ sector and its effects on the pNGB potential, as well as elucidating the possibility of co-generating the Higgs and $\Sigma$ sector scales. However, because the $\Sigma$ sector must be near strong coupling and its interactions with the Higgs sector can be a strong perturbation, there may be important higher-order effects neglected in this description. In the following section, we give an effective description of strongly-coupled UV completions and argue that the qualitative features remain the same.

\subsection{Strongly-Coupled Auxiliary Sectors}
\label{sec:strongcoupling}

We now focus on the case that $H$ and $\Sigma$ emerge from independent strongly-coupled sectors with compositeness scales $\Lambda_\Sigma < \Lambda_H$.

\subsubsection{Composite Higgs --- $SO(5)/SO(4)$}

The global symmetries  of the two sectors are $SO(5)_H$ and $SO(4)_\Sigma$. At scales above $\Lambda_H$, the two sectors are weakly coupled by an operator explicitly breaking $SO(5)_H \times SO(4)_\Sigma \rightarrow SO(4)$,
\beq
\label{eq:kappaUV}
\mathcal{L}\supset\mathcal{O}_{\kappa^2}=\kappa^2_{I j}\mathcal{O}_H^I \mathcal{O}_\Sigma^j
\eeq
The spurion $\kappa^2_{I j}$ parameterizes the breaking,
\beq
\kappa^2_{I j} = \kappa^2 \begin{pmatrix} 1 & 0 & 0 & 0 \\ 0 & 1 & 0 & 0 \\ 0 & 0 & 1 & 0 \\ 0 & 0 & 0 & 1 \\ 0 & 0 & 0 & 0 \end{pmatrix}.
\eeq
We normalize these operators so that, in terms of the low-energy goldstone fields, $\mathcal{O}_H^I = H^I (1 + ...)$ and $\mathcal{O}_\Sigma^j = \Sigma^j (1 + ...)$. The neglected terms are higher derivative in the goldstone fields. A convenient realization of the pNGB manifold is given by
\begin{align}
H_I & = \frac{f_H}{\sqrt{2}} e^{i \Pi^a_{h} T_a/v_H}  \begin{pmatrix} 0 \\ 0 \\ 0 \\ s_h \\ 0 \end{pmatrix} + \frac{f_H}{\sqrt{2}}\begin{pmatrix} 0 \\ 0 \\ 0 \\ 0 \\ c_h \end{pmatrix}, \\
\Sigma_I & = \frac{f_\Sigma}{\sqrt{2}} e^{i \Pi^a_{\Sigma} T_a/f_\Sigma}  \begin{pmatrix} 0 \\ 0 \\ 0 \\ 1 \end{pmatrix}.
\end{align}
The fields $\Pi^a_h$ and $\Pi^a_\Sigma$ correspond to the pNGBs of the broken $SU(2)_H$ and  $SU(2)_\Sigma$, with a linear combination 
\be 
\Pi^a_G=\frac{v_H}{v}\Pi^a_h +\frac{f_\Sigma}{v} \Pi^a_h
\ee 
absorbed by the gauge bosons and the remaining
\be
\Pi^a_A=\frac{f_\Sigma}{v}\Pi^a_h -\frac{v_H}{v}\Pi^a_\Sigma
\ee 
obtaining a mass from the explicit  breaking.

When $\mathcal{O}_{\kappa^2}$ is a sufficiently weak perturbation on both the $\Sigma$ and $H$ sectors, the leading effect in the IR at scales below $\Lambda_H$ is to generate the tadpole term $\kappa^2 H \cdot \Sigma$ in the pNGB potential. In the parameter space of interest $\mathcal{O}_{\kappa^2}$ will always be a weak perturbation on the fundamental $H$ sector at $\Lambda_H$, but it may not be a weak perturbation on the $\Sigma$ sector;
by NDA~\cite{Weinberg:1978kz,Manohar:1983md,Georgi:1986kr} $\mathcal{O}_{\kappa^2}$ can be a strong perturbation on the $\Sigma$ sector if $\kappa^2 v_H \gsim \Lambda^2_\Sigma f_\Sigma$, as for the linear sigma model above.

The effects on the pNGB Higgs potential are determined by treating $H$ as a background field and integrating out the $\Sigma$ sector at $\Lambda_\Sigma$ to obtain the full Goldstone potential,
\be
\label{eq:Vkappa}
V_{\kappa^2}(h, \Pi_A) \equiv \hat V\left(\frac{\Sigma^j}{f_\Sigma}, \frac{\kappa^2_{Ij} H^I}{\Lambda^2_\Sigma f_\Sigma}\right)\Lambda^2_\Sigma f^2_\Sigma,
\ee
with $\hat V$ a function with $\mathcal O(1)$ coefficients.
Terms of the form $\kappa^2_{I j} \Sigma^j H^I$ generate a potential for both $h$ and $\Pi_A$, while the invariant $\kappa^2_{K j}\kappa^2_{I j} H^K H^I$ generates a potential only for $h$. For the Higgs potential, we obtain simply
\be
\label{eq:Vkappa}
V_{\kappa^2}(h) = \Lambda_\Sigma^2 f_\Sigma^2 \hat V\left( \frac{\kappa^2 f_H}{\Lambda^2_\Sigma f_\Sigma} s_h\right).
\ee
This term fully describes the IR contributions from the $\Sigma$ sector, and connects the size of the tadpole to the higher-order terms. 
For instance, these terms will generate a contribution to $\alpha$,
\be
\delta \alpha \simeq {\cal O}\left(\frac{\kappa^4}{\Lambda_\Sigma^2 f_\Sigma^2}\right),
\ee
again consistent with the results for a linearly-realized auxiliary sector, although with undetermined coefficient. Higher-order terms in \Eref{eq:Vkappa} can also give $\mathcal{O}(1)$ shifts in the masses of the extra Higgs sector states $\Pi_A$. For example, the tadpole and first leading contribution to the masses of the $\Pi_A$ have the form
\begin{flalign}
\label{eq:Amass}
V(\Pi_A) & = \kappa^2\Sigma\cdot H\left(1 + c\frac{\kappa^2 \Sigma\cdot H}{\lambda_\Sigma^2 f^2_\Sigma}\right) + \hc  \\
& \simeq \frac{1}{2}\kappa^2 \frac{v_H}{f_\Sigma}\left (1 + 2c\frac{\kappa^2 v_H}{\Lambda^2_\Sigma f_\Sigma}\right) \Pi_A^2 + \ldots \nonumber
\end{flalign}
Integrating out the $\Sigma$ sector also generates terms of the form $f\left(\frac{\kappa^2 v_H}{\Lambda^2_\Sigma f_\Sigma}\right)|D_\mu \langle \Sigma \rangle|^2$, which effectively shift the auxiliary EWSB vev from $f_\Sigma$ by an amount parametrically of the same size as the back-reaction in the linear sigma model, \Eref{eq:sigmavevshift}.

For $\Lambda_\Sigma \simeq 4 \pi f_\Sigma$, back-reaction and higher-order terms result in $\lsim {\cal O}(1)$ shifts to $f_\Sigma$ and $\alpha$, analogous to the results of the preceding subsection.  

\subsubsection{Twin Higgs --- $SO(8)/SO(7)$}
\label{sssec:thaux}

In the Twin model, the $H$ sector has an $SO(8)_H$ global symmetry and the $\Sigma$ sector has a custodial $SO(4)_{\Sigma_A} \times SO(4)_{\Sigma_B}\times {\mathbb Z}_2$ global symmetry. The coupling of the $H$ and $\Sigma$ sector extends the form of the $SO(5)/SO(4)$ model, explicitly breaking the global symmetry to $SO(4)_A \times SO(4)_B \times {\mathbb Z}_2$,
\be
\label{eq:kappaUV}
\mathcal{L}\supset\mathcal{O}_{\kappa^2}=\hat\kappa^2_{(A)I j}\mathcal{O}_H^I \mathcal{O}_{\Sigma_A}^j  +\hat\kappa^2_{(B)I j}\mathcal{O}_H^I \mathcal{O}_{\Sigma_B}^j
\ee
Following the same analysis as for the $SO(5)/SO(4)$ model, the IR contribution to the Higgs potential has the form
\be
\label{eq:kappaIRTwin}
V_{\kappa^2}(h) = \Lambda^2_\Sigma f^2_\Sigma  \left[ \hat{V}\left( \frac{\kappa^2 f_H}{\Lambda^2_\Sigma f_\Sigma} s_h\right) + \hat{V}\left( \frac{\kappa^2 f_H}{\Lambda^2_\Sigma f_\Sigma} c_h\right) \right],
\ee
with the structure enforced by the ${\mathbb Z}_2$ symmetry. We choose to express the potential in terms of a redefined parameter $\kappa^2 \sim \hat\kappa^2$ to normalize the tadpole term as $\kappa^2_{Ij} H_A^I \Sigma_A^j$.

As for the composite example, the higher-order terms are parametrically the same size as calculated in the linear realization for $\Lambda_\Sigma \simeq 4 \pi f_\Sigma$, such that the tadpole due to $f_H$ can readily constitute a significant perturbation on the $\Sigma_B$ sector.  In addition, we expect the operators in the pNGB potential to be generated with ${\cal O}(1)$ coefficients, permitting the possibility that these terms can generate additional positive contributions to $\alpha$, perhaps alleviating the need for additional UV contributions required to overcome the $\delta \alpha < 0$ from the SM top sector.

Another notable detail is that non-negligible higher-order terms coupling $H$ and $\Sigma$ should be generated.  
Depending on their sign and size, these terms may lead to complete breaking of $SU(2)_B \times U(1)_B$ (in the event that Twin hypercharge is gauged).  In particular, as $f_H \gg v_H$ and $f_{\Sigma_B} \sim f_{\Sigma_A}$, higher-order terms can drive $SU(2)$ alignment of $\vev{H_A}$ and $\vev{\Sigma_A}$ but misalignment of $\vev{H_B}$ and $\vev{\Sigma_B}$ even with $\mathbb{Z}_2$-symmetry. In this case, $SU(2)_B \times U(1)_B$ is fully broken while $SU(2)_A \times U(1)_A \rightarrow U(1)_{\rm EM}$, avoiding a massless Twin hypercharge boson.

Finally, in the Twin case, there is the additional question of the origin of the two auxiliary sectors.
$\Sigma_A$ and $\Sigma_B$ may be part of a single strongly-interacting gauge sector $\mathcal{G}$ or two independent strongly-interacting sectors $\mathcal{G}_A$ and $\mathcal{G}_B$ related by the ${\mathbb Z}_2$.
The former naturally admits the appealing ``waterfall'' of induced breakings described above. In the linear sigma model, this case corresponds to a large coupling between the $A$ and $B$ sectors, $\lambda_\Sigma \sim \delta_\Sigma$---analogously, for a single strongly-coupled sector, we expect sizable couplings between $\Sigma_A$ and $\Sigma_B$. The condensation in the Higgs sector at $\Lambda_H$ generates a scale in the $B$ auxiliary sector, triggering its condensation. For example, the $\Sigma$ sector could be a conformal technicolor-like sector near a strongly-coupled fixed point at $\Lambda_H$, with some techniquarks $\mathcal O_\Sigma \sim \psi_\Sigma \bar\psi_\Sigma$ acquiring $SU(2)_A$-preserving masses proportional to $f_H$. This triggers a chiral symmetry-breaking phase for both the $A$ and $B$ sectors, which in turn generates the tadpole for $H_A$, inducing EWSB. In this scenario, the scale in the $A$ auxiliary sector is directly related to the scale in the $B$ auxiliary sector so we expect $f_{\Sigma_B} \sim f_{\Sigma_A}$ in the absence of tuning. The scales of the Higgs and $\Sigma_A$ sector are therefore directly connected as $\Lambda^3_{\Sigma_A}\sim \kappa^2 f_H$, and the viable parameter space $f_{\Sigma_A}\sim 50-70\GeV$ requires $f_H \sim \TeV$. Alternatively, if $\mathcal{G_A}$ and $\mathcal{G_B}$ are independent, the Twin sector can induce $f_{\Sigma_B} \gg f_{\Sigma_A}$ which can increase the size of the extra contributions to the Higgs potential.

\subsection{UV Considerations}
\label{subsec:uvcompletion}

Finally, we highlight some of the additional important issues that should be addressed by UV completions attempting to explain the origin of the Higgs and $\Sigma$ sectors.

As stressed throughout, the mechanism of induced EWSB requires $\alpha > 0$.  
This does not present a particular challenge in the MCHM, as gauge contributions to $\alpha$ are positive and UV-sensitive, so can easily be arranged to give $\alpha > 0$ if the gauge partners are heavier than the top partners. 
Depending on the structure of the $\Sigma$ sector, however, there can also be UV contributions to the Higgs potential $\propto \kappa^2$, which may need to be suppressed to avoid tuning.
For example, $\Sigma$ can emerge from an asymptotically free technicolor-like sector that is weakly coupled at the scale $\Lambda_H$ with $O_\Sigma$ formed from elementary fermions, $\mathcal{O}_{\Sigma} = \Sigma^I + \ldots \simeq \frac{1}{f_\Sigma \Lambda_\Sigma}\psi_\Sigma \bar \psi_\Sigma$. Contributions to the potential for $H$ are cut off at $\Lambda_H^2$ and give a leading one-loop UV contribution
\be
\label{eq:kappaUV}
V_{\kappa^2,UV} \sim \frac{\Lambda_H^2}{16\pi^2} \left(\frac{\kappa^2_{I j}}{\Lambda_\Sigma f_\Sigma} H^I\right)^2 \simeq \Lambda_\Sigma^2 f_H^2 \left(\frac{\kappa^2 f_H}{\Lambda^2_\Sigma f_\Sigma}\right)^2 s_h^2.
\ee
This exceeds the IR-generated quadratic term by a factor $\propto \frac{f_H^2}{f_\Sigma^2}$, so could dominate over the radiative top sector tuning if unsuppressed. The UV $s_h^4$ term is of comparable size to the IR-generated term, and higher-order UV terms are subdominant. 
A more general $\Sigma$ sector can entirely avoid such overly-large UV contributions to the potential if the scaling dimension is $[\mathcal{O}_\Sigma] \leq 2$ at $\Lambda_H$. This is trivially satisfied in a scalar linear $\Sigma$ model, or can occur in a conformal technicolor-like theory near a strongly-interacting fixed point with large anomalous dimension for the fermion bilinears.

In the Composite Twin Higgs case, the ${\mathbb Z}_2$ removes the quadratic UV-sensitivity of the top, gauge and auxiliary sector contributions to $\alpha$, potentially making it more difficult to realize $\alpha > 0$.
The top and gauge sectors generate $\delta \alpha < 0$ in the IR, making additional positive IR contributions from the $\Sigma$ sector even more desirable. 
Fortunately, as seen in the linear sigma model of \Sref{sec:linearsig}, these contributions can be enhanced due to the presence of the extra $B$ sector with $f_{\Sigma_B} \gsim f_{\Sigma_A}$, and can readily be of comparable size to the experimentally-required value $\alpha_{\rm obs}$. Because the top sector radiative contribution can also be ${\cal O}(\alpha_{\rm obs})$ (see \Sref{subsec:twintoptuning}), this means $\Sigma$ sector IR contributions are in principle sufficient to generate $\alpha = \alpha_{\rm obs} > 0$.

Even if IR contributions to $\alpha$ from the $\Sigma$ sector are insufficient, though, other possible sources for $\alpha > 0$ exist. These include embedding the $\tau$ or bottom sector in a larger representation of $SO(8)$, which can give a UV contribution to $\alpha$ of the right sign and size to tune against the top contribution, analogous to the mechanism for increasing $\beta$ in $SO(5)/SO(4)$ models \cite{Carmona:2014iwa}. In Twin Higgs models based on an $SU(4)/SU(3)$ coset, overly-large contributions from the gauge sector, $\delta \alpha \sim g^2$, may be a concern~\cite{Barbieri:2015lqa}, but such contributions are forbidden if the global symmetry is expanded to $SO(8)$ \cite{Chacko:2005un,Barbieri:2015lqa}. This indicates small explicit breakings of $SO(8)$ to $SU(4)$ may also be useful to obtain $\alpha > 0$. 

A UV completion should also address the potentially dissatisfying coincidence of scales, $f_\Sigma \lsim m_h \sim v$.
In the context of SUSY, for EWSB induced by a strongly-coupled $\Sigma$ sector, \Rref{Galloway:2013dma} suggested that the auxiliary sector could be near a strongly-coupled superconformal fixed point in the UV. Then, SUSY breaking triggers confinement at a scale close to that of the scalar soft masses. As alluded to above for the Twin Higgs, one could imagine a similar mechanism here, namely that confinement in the nearly-conformal auxiliary sector is triggered by breaking of the approximate global symmetry at $\Lambda_H$ (though, admittedly, there are more known examples of superconformal theories). A similar scenario can be realized in the $SO(5)/SO(4)$ model if an additional operator of comparable strength to $\mathcal{O}_{\kappa^2}$ couples $\mathcal{O}_H$ to an $SO(4)$ singlet in the $\Sigma$ sector. Regardless of the solution, it must avoid introducing a hierarchy problem in the $\Sigma$ sector, which would of course spoil the improved naturalness exhibited by these models.

\section{Experimental Constraints}
\label{sec:expt}

Induced EWSB is subject to both indirect constraints from measurements of Higgs properties and electroweak precision tests, and direct constraints from searches for additional states associated with the auxiliary sector.
These constraints have been extensively studied in \cite{Chang:2014ida}, with emphasis on phenomenological models and applications to supersymmetry.
Notably, there exists a tension between electroweak precision tests and direct searches for vector resonances, which favor larger values of $f_\Sigma$, and Higgs measurements (both of Higgs properties and searches for extended Higgs sector states), which favor smaller values of $f_\Sigma$.
Here, we summarize these results, and highlight some of the main differences in the MCHM or Twin scenario.

The presence of an additional source of EWSB modifies Higgs couplings to SM states.  If the auxiliary sector is strongly-coupled, this results in a universal enhancement of Higgs couplings to fermions and a suppression of couplings to gauge bosons, parameterized by the ratios
\begin{align}
\label{eq:kV}
\kappa_f & \equiv \frac{g_{h f \bar{f}}}{g_{h f \bar{f}}^{\rm(SM)}} = \frac{1}{\sqrt{1 - \frac{f_{\Sigma_{(A)}}^2}{v^2}}}, \\
\label{eq:kf}
\kappa_V & \equiv \frac{g_{h V V}}{g_{h V V}^{\rm(SM)}} = \sqrt{1 - \frac{f_{\Sigma_{(A)}}^2}{v^2}}.
\end{align}
The allowed values of $f_\Sigma$ are thus constrained by the combined ATLAS and CMS Higgs measurements \cite{ATLAS-CONF-2015-044,Khachatryan:2014jba,Aad:2015gba}---for a strongly-coupled auxiliary sector, $f_\Sigma \lsim 0.3 v$ \cite{Chang:2014ida}.
Motivated by the discussion of \Sref{sec:SigmaDynamics}, we focus on strongly-coupled auxiliary sectors here.
However we do note that, if the auxiliary sector is at least somewhat weakly-coupled, the constraints vary due to the mixing between the Higgs and the radial mode of the auxiliary sector.  This mode couples to gauge bosons but not to fermions, so mixing partially restores the depletion of $\kappa_V$ while also reducing the enhancement of $\kappa_f$.

In pNGB Higgs models, there is additional universal suppression of Higgs couplings due to $\frac{v_H^2}{f_H^2}$ corrections,
\be
\kappa^{\rm(pNGB)}_h \simeq \sqrt{1 - \frac{v_H^2}{f_H^2}}.
\ee
While this counteracts the enhancement of Higgs coupling to fermions, it also further suppresses coupling to vector bosons. Since current measurements favor a slight enhancement $\kappa_V = 1.05 > 1$, constraints on $f_\Sigma$ can be somewhat more stringent for smaller values of $f_H$ than in the usual induced EWSB scenario described above.

For Twin Higgs models, there is further additional suppression of Higgs couplings to visible SM states due to decays to Twin sector 
states~\cite{Burdman:2014zta}.  For instance, supposing the couplings to SM and Twin $b$ quarks respect the $\mathbb{Z}_2$, the Higgs is expected to decay to Twin $b$'s with width $\Gamma_{h \rightarrow b' \bar{b}'} \simeq \frac{v_H^2}{f_H^2} \Gamma_{h \rightarrow b \bar{b}}$, leading to a suppression factor
\be
\kappa^{\rm(TH)}_h \simeq \frac{1}{\sqrt{1 + \frac{v_H^2}{f_H^2} {\rm Br}^{\rm(SM)}(h \rightarrow b \bar{b})}}
\ee
where ${\rm Br}^{\rm(SM)}(h \rightarrow b \bar{b}) = 0.577$ for $m_h = 125 \GeV$.  However, depending on the exact details of the quark couplings, this decay may be suppressed and a variety of Higgs decays to Twin sector states, including displaced decays, may be possible (see, \eg, \cite{Craig:2015pha, Curtin:2015fna, Csaki:2015fba}).

\begin{figure}
\centering
\includegraphics[width=.9\columnwidth]{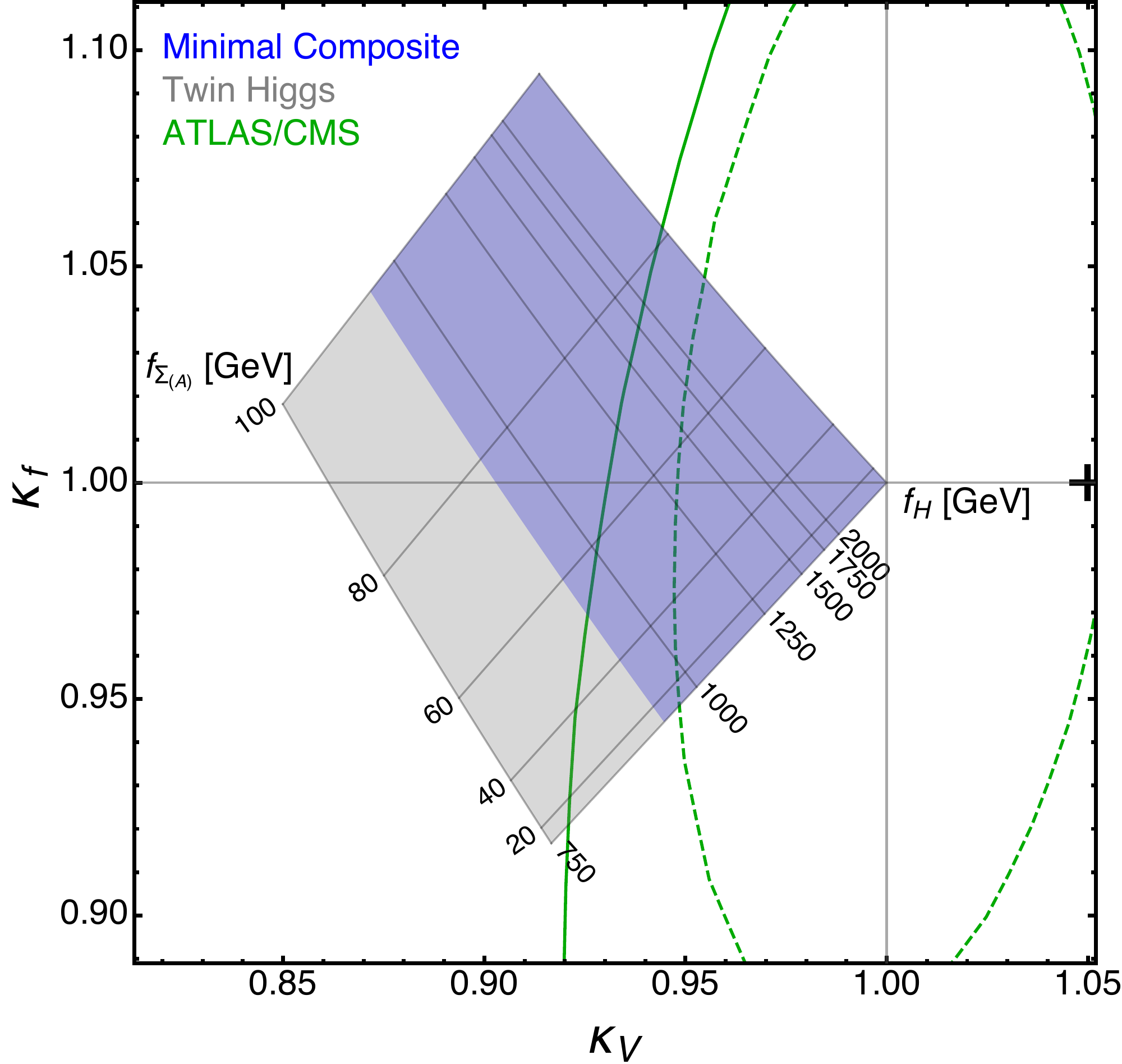}
\caption{
\label{fig:kappaVkappaf}
Values of $(\kappa_V,\kappa_f)$ in Minimal Composite (blue) and Twin (gray) models for $0 \leq f_\Sigma \leq 100 \GeV$ and $750 \GeV \leq f_H$. Contours correspond to values of $f_\Sigma$ and $f_H$ in a Twin Higgs model with unsuppressed decays to Twin $b$'s. Solid elliptical contour corresponds to the combination of current ATLAS and CMS measurements, with central value $(\kappa_V,\kappa_f) = (1.05,1)$~\cite{ATLAS-CONF-2015-044}.
Dashed contour corresponds to projections from~\cite{Chang:2014ida} assuming central value $(\kappa_V,\kappa_f) = (1,1)$.
}
\end{figure}

In \Fref{fig:kappaVkappaf}, we plot the $(\kappa_V,\kappa_f)$ that can occur in induced EWSB models with a pNGB Higgs and a strongly-coupled auxiliary sector, as well as the combined ATLAS and CMS measurements \cite{ATLAS-CONF-2015-044}.
We consider both a general MCHM model (\ie, with additional suppression $\kappa_h^{\rm(pNGB)}$ relative to \Erefs{eq:kV}{eq:kf} only), as well as a TH model with unsuppressed decays to Twin $b$'s (with additional suppression $\kappa_h^{\rm(pNGB)} \kappa_h^{\rm(TH)}$).
We also show projected limits from \cite{Chang:2014ida} assuming $\sqrt{s} = 14\TeV, {\cal L} = 300 \text{ fb}^{-1}$ and central value $(\kappa_V,\kappa_f) = (1,1)$.

There are also constraints from direct searches for states associated with the auxiliary sector, which generally require these states to be at least somewhat heavy.
First, there are the additional Higgs sector states due to the presence of a second electroweak doublet $\Sigma_{(A)}$.
These states have masses related to the size of the $H \cdot \Sigma$ terms connecting the two sectors as, in the limit such terms vanish, the Higgs and $\Sigma$ sectors exhibit separate $SU(2)$ symmetries. Taking $\kappa^2$ to be the only such term and neglecting non-quadratic Higgs terms, one finds \cite{Chang:2014ida}
\be
\kappa_0^2 = \frac{m_h^2 v_H}{f_\Sigma}
\ee
and, correspondingly,
\be
\label{eq:mA}
m_{A,0}^2 = m_{H^\pm}^2 \simeq m_h^2 \frac{v^2}{f_\Sigma^2}.
\ee
In our case this relationship is modified by higher-order terms. First, in the pNGB Higgs potential, $\alpha > 0$ yields a negative quartic, which would tend to enhance $m_A$ relative to the above estimate, but we also expect higher-order terms including $\beta \neq 0$ to be generated. 
For instance, the natural $5+1$ top sectors considered in \Sref{subsec:5plus1} tend to generate (positive) $\beta \simeq \frac{\beta_{\rm SM}}{2}$, which would decrease $m_A$, while for Twin Higgs models unbroken ${\mathbb Z}_2$ enforces $\beta = - \alpha$, further enhancing $m_A$.
Second, for a strongly-coupled auxiliary sector, higher-order $H \cdot \Sigma$ terms can yield $\mathcal{O}(1)$ corrections, as in \Eref{eq:Amass}. These two effects represent a `theoretical uncertainty' in the relation between $(f_H,f_\Sigma)$ and the mass of Higgs resonances in the auxiliary sector.

\begin{figure}
\centering
\includegraphics[width=0.9\columnwidth]{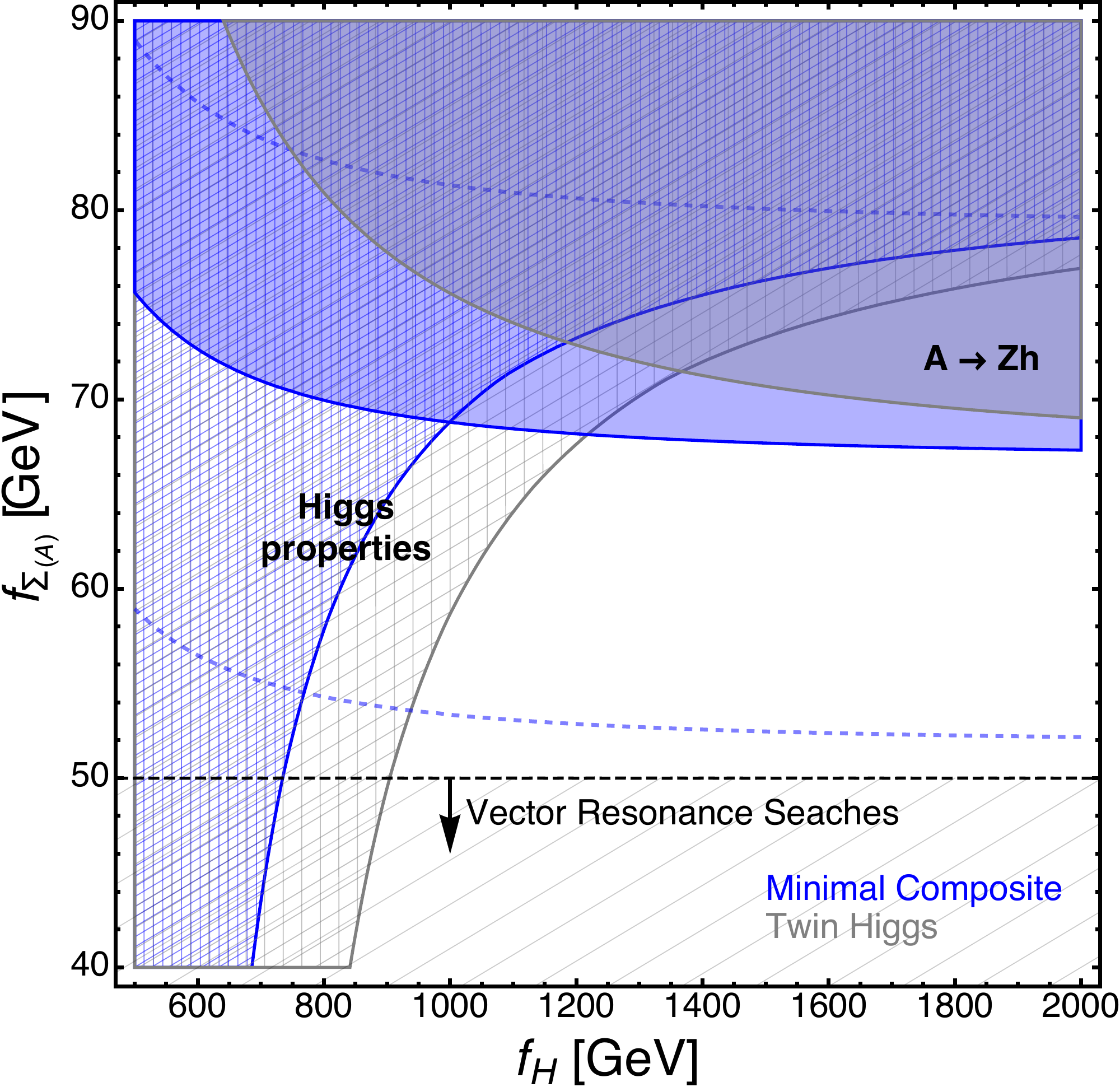}
\caption{
\label{fig:ffSigma}
Regions of $(f_H,f_\Sigma)$ excluded by Higgs coupling measurements (hatched) and direct $A \rightarrow Zh$ searches (solid) for Minimal Composite (blue) and Twin (gray) Higgs models.
Solid regions correspond to $m_A$ with $\beta=0$ for MCHM and $\beta = \alpha$ for Twin Higgs, see text for details. 
Dashed blue contours represent the effect of rescaling $m_A^2$ by $0.6$ (lower) or $1.4$ (upper) and thus represent the theoretical uncertainty on the solid blue line.
The dashed black line denotes approximate lower bound $f_\Sigma \gsim 50 \GeV$ from vector resonance searches.}
\end{figure}

Direct searches for heavy Higgs bosons constrain $m_A$, with the dominant constraint in much of the parameter space coming from the CMS search for $A \rightarrow Z h \rightarrow \ell^+ \ell^- b \bar{b}$ \cite{Khachatryan:2015lba}, which requires $m_A \gsim 460 \GeV$~\cite{Chang:2014ida}. 
In \Fref{fig:ffSigma}, we show approximate constraints from $A \rightarrow Zh$ and Higgs property measurements in the $(f_H,f_\Sigma)$ plane, again supposing a strongly-coupled auxiliary sector.
To determine $m_A$, we rescale $m_{A,0}$ by $\frac{\kappa}{\kappa_0}$ where $\kappa$ is determined from \Eref{eq:simpletadpoleV} for the Minimal Composite model (\ie, neglecting $\beta$) and from \Eref{eq:tadpolePotentialTH} with $\beta = \alpha$ for the Twin Higgs model (\ie, neglecting ${\mathbb Z}_2$ breaking).
In both cases, this corresponds to neglecting corrections due to higher-order $H \cdot \Sigma$ terms.
We have tested the approximate size of the corrections mentioned above in several specific cases;
to capture and summarize the potential importance of the neglected effects, we also show the impact of rescaling $m_A^2 \rightarrow (0.6,1.4) m_A^2$ for the MCHM. An uncertainty band of similar proportion also applies for the Twin Higgs.

A second set of constraints comes from vector resonances. If the auxiliary sector is indeed strongly-coupled, we expect vector resonances with masses $m_\rho \sim 4 \pi f_\Sigma$ associated with the strong dynamics \cite{Carone:1993vg}.  These ``technirhos'' are constrained both by direct searches (notably, $\rho^\pm \rightarrow W^\pm Z$ \cite{Aad:2014pha}) and by electroweak precision measurements \cite{Carone:1992rh}.  The exact constraints depend on the properties of the technirhos, which depend on the details of the unknown strong dynamics.  However, for lighter technirhos (such as those predicted by a QCD-like auxiliary sector), these can be the dominant constraints, eliminating the majority of the allowed parameter space \cite{Chang:2014ida}.  Thus, for a truly strongly-coupled auxiliary sector, the strong dynamics must be such that the vector resonances are at least somewhat heavy.
For instance, the (non-excluded) strongly-coupled benchmarks considered in \cite{Chang:2014ida} would require $f_\Sigma \gsim 50-55 \GeV$.
Meanwhile, perturbativity generally places an upper bound on $m_\rho$.

Finally, for pNGB Higgs models, top partner searches are of course also relevant. There are a variety of searches focusing on a minimal charge-$2/3$ top partner decaying via $T \rightarrow b W, t Z, th$, which currently require $m_T \gsim 700 \GeV$ \cite{Aad:2015kqa,Khachatryan:2015oba,Aad:2016qpo}. A top partner of this variety is expected to be somewhat light as it is responsible for cutting off quadratic divergences due to the SM top quark. However, in `maximally natural' models, the full global symmetry is likely restored not too far above $m_T$ (see \Sref{sec:radiativetuning}). As a result, searches for other states implied by the global symmetry, such as heavy charge-$1/3$ $B$-quarks \cite{Aad:2015mba,Khachatryan:2015gza} or exotic charge-$5/3$ quarks \cite{Chatrchyan:2013wfa,Aad:2015mba} (present in complete multiplets of custodial $SO(4)$) may also be relevant \cite{Agashe:2006at,Kearney:2013cca}. In particular, for Twin Higgs models, the lightest top partner responsible for regulating the quadratic divergences is uncolored, leading to weak constraints from the LHC. But natural models likely exhibit colored top partners not too much heavier than the uncolored twin top (as in \Sref{subsec:twintoptuning}), which may be probed up to $m_\ast \sim 2.5 \TeV$ at the LHC~\cite{Cheng:2015buv}.

\section{Conclusion}
\label{sec:conclusion}

Tadpole-induced electroweak symmetry breaking gives an alternative structure for the low-energy potential of a pNGB Higgs model. This structure allows the desired EWSB pattern with $m_h=125\GeV$ and $v_H \ll f_H$ to be achieved in Composite Higgs models that could not otherwise realize a large enough quartic term $\beta$ without excessive tuning. Unlike other tree-level modifications to the pNGB Higgs potential, which focus on increasing the quartic term $\beta$ (\eg, Little Higgs), the tadpole structure simply makes $\beta$ irrelevant in the limit $v_H \ll f_H$. 

In $SO(5)/SO(4)$ minimal composite Higgs models (MCHM), this mechanism makes viable the minimal representations of the 3rd generation partners (as in \Mfive). The resulting tuning is comparable to a purely radiative potential generated by 3rd generation partners in larger representations. In the case of the Twin Higgs, the radiative contributions from the minimal representations of the top sector can be made substantially smaller, and the mechanism of induced EWSB allows the tuning to be reduced by a factor of $\sim5$ compared to the `irreducible' $\frac{f_H^2}{2 v_H^2}$ tuning of a purely radiative potential. This allows a fully natural pNGB potential with $f_H \simeq 1 \TeV$ and colored top partners at $\sim 2 \TeV$. The tadpole mechanism in the Twin Higgs model also has the advantage of incorporating spontaneous ${\mathbb Z}_2$ breaking and full breaking of the mirror $U(1)_{\rm EM,B}$.

While these pNGB Higgs models share many features in common with supersymmetric models of induced~EWSB~\cite{Samuel:1990dq,Dine:1990jd,kagantalk,Azatov:2011ht,Azatov:2011ps,Galloway:2013dma,Chang:2014ida}, 
there are interesting differences. First, in SUSY models, both the $H$ and $\Sigma$ sectors inherit their scale from an external SUSY breaking sector, while in the composite pNGB case the scale $f_H$ can directly trigger $f_\Sigma$. Second, although in both cases the striking phenomenology is in the Higgs sector, Higgs compositeness generates additional deviations in Higgs properties not present in SUSY. If the fermionic top partners of the pNGB Higgs model are within reach, their signatures also differ substantially from the signatures of the scalar stops in SUSY models.

In the most appealing version of the model, the scale $f_H$ of global symmetry breaking triggers a waterfall of breaking where $f_H$ dynamically induces the smaller $f_\Sigma$ which in turn induces $v_H$, naturally connecting the scales $f_\Sigma \ll v_H \ll f_H$.
In this scenario, the compositeness scale must be $f_H\sim\TeV$.
Meanwhile, the combination of Higgs property measurements and searches for the new auxiliary sector states set both upper and lower bounds on the scale $f_\Sigma$, and it is non-trivial that there is consistent parameter space for this model with new TeV-scale physics.

In tadpole-induced pNGB Higgs models, a wealth of interesting phenomenology from both the $\Sigma$ sector and Higgs compositeness may be within reach of the LHC. The plethora of signals could include modifications of Higgs properties due to both compositeness and the auxiliary EWSB component, extra charged and pseudoscalar Higgs states, auxiliary sector vector resonances lighter than $1\TeV$, and colored composite top partners at $\gsim$ TeV.  In the case of the Twin Higgs, further consequences of the mirror sector, including invisible and/or exotic Higgs decays, may be observable. It has not escaped our attention that the auxiliary sector generically contains composite singlet pseudoscalars at the scale $\Lambda_\Sigma\sim 4\pi f_\Sigma \sim 750\GeV$ with large branching ratios to diphotons \cite{Harigaya:2016pnu,Harigaya:2015ezk,Nakai:2015ptz,Molinaro:2015cwg,Franzosi:2016wtl,Chiang:2015tqz,Bai:2015nbs}, which may be able to explain recent hints for a resonance at LHC13 \cite{CMS-PAS-EXO-12-045,ATLAS-CONF-2015-081}. In particular, small mixings between the auxiliary sector and singlet pseudoscalars in the composite Higgs sector \cite{No:2015bsn,Low:2015qep,Belyaev:2015hgo,Bellazzini:2015nxw} can lead to an appreciable gluon fusion production cross section even if the auxiliary sector contains no colored states.

Not only can tadpole-induced models feature a pNGB potential with a fully natural scale for EWSB, but in fact searches at LHC13 and future colliders will likely be able to probe the entire remaining range of viable models independent of any naturalness arguments.

\vspace{0.5cm}

\paragraph*{Acknowledgements:} We thank Nathaniel Craig for useful discussions at various stages of this work and also wish to congratulate him for his productivity. We also thank Zackaria Chacko, Roberto Contino, Juliano Panico, Ennio Salvioni, Yuhsin Tsai, and Andrea Tesi. We thank Alex Kagan and Adam Martin for friendly coordination regarding their upcoming related paper~\cite{AdamAlex}. Fermilab is operated by Fermi Research Alliance, LLC under Contract No. \protect{DE-AC02-07CH11359} with the United States Department of Energy.

\begin{widetext}
\appendix
\section{Expressions for two-site models}
\label{app:radiative}

\subsection{5+1}

The Lagrangian for the two-site model defining the mass mixings and Yukawa couplings of the 5+1 top sector is
\begin{align*}
 \mathcal{L} = & - m_1 (\psi_{1} \psi^c_{1})
   - m_4 (\psi_{4}^i {\psi_{4}^c}^i)
   - m_R (\psi_{1} t_{R}) 
  - \frac{y_L f_H}{\sqrt{2}}\left(t_{L}({\psi^c_{4}}^{(2)} + \cos\frac{h}{f_H}{\psi^c_{7}}^{(4)}) + \sin\frac{h}{f_H} t_{L}\psi^c_{1}\right) + \hc
\end{align*}

The breaking of the global symmetry due to the top sector mixings is completely parameterized by $y_L$.  The mixing angles are $\sin^2\theta_R=\frac{m^2_R}{m^2_R + m_1^2}$ and $\sin^2\theta_L=\frac{y_L^2 f_H^2}{y_L^2 f_H^2 + m_4^2}$, and the top partners mixing with the elementary sector obtain masses $M_1^2 = m_1^2 + m_R^2$ and $M_4^2 = m_4^2 + y_L^2 f_H^2$.
The Top Yukawa coupling is given to leading order in $\sin\frac{h}{f_H}$ as $y_t = \frac{m_4}{f}\sin\theta_R\sin\theta_L $. 

The mass matrix can be diagonalized perturbatively in $\sin\frac{h}{f_H}$, giving a Coleman-Weinberg contribution to the effective potential,
\begin{equation}
\Delta\alpha = -\frac{3 y_L^2}{16\pi^2 f^2}\left(
m_4^2\left(1+\log\frac{\mu^2}{M_4^2}\right)
-m_1^2\left(1+\log\frac{\mu^2}{M_1^2}\right) 
+\frac{m_1^2 y_L^2 f_H^2}{M_4^2-M_1^2}\log\frac{M_4^2}{M_1^2} \right).
\end{equation}
$\Delta\beta$ is obtained in the same fashion.

\subsection{8+1}
The Lagrangian for the two-site model defining the mass mixings and Yukawa couplings of the 8+1 top sector generalizes the 5+1 model to the twin case by extending the coset to $SO(8)/SO(7)$ and including B-sector elementary and composite tops,
\begin{align*}
 \mathcal{L} = & - m_1 (\psi_{1,A} \psi^c_{1,A}+\psi_{1,B} \psi^c_{1,B}) 
   - m_7 (\psi_{7,A}^i {\psi_{7,A}^c}^i+\psi_{7,B}^i {\psi_{7,B}^c}^i) 
   - m_R (\psi_{1,A} t_{R,A}+\psi_{1,B} t_{R,B}) \\  
 & - \frac{y_L f_H}{\sqrt{2}}\left(t_{L,A}({\psi^c_{7,A}}^{(2)} + \cos\frac{h}{f_H}{\psi^c_{7,A}}^{(4)}) + \sin\frac{h}{f_H} t_{L,A}\psi^c_{1,A}\right) \\
 & - \frac{y_L f_H}{\sqrt{2}}\left(t_{L,B}({\psi^c_{7,B}}^{(6)} + \cos\frac{h}{f_H}{\psi^c_{1,B}}) + \sin\frac{h}{f_H} t_{L,B}{\psi^c_{7,B}}^{(4)}\right)  + \hc
\end{align*} 

It is simplest to proceed directly to the radiative potential following Ref.~\cite{Low:2015nqa}. We obtain $\alpha$ by expanding to order $\sin^2\frac{h}{f}$ (Ref.~\cite{Low:2015nqa} makes a similar expansion in $y^2$ instead of $\sin^2\frac{h}{f}$),
\begin{equation}
\begin{split}
\Delta\alpha = &  -\int\frac{d^4p}{(2\pi)^4}\frac{y_L^4 \left({m_1}^2 p^2+{m_7}^2 \left({m_R}^2-p^2\right)\right)^2}{p^2 \left(-{m_1}^2-{m_R}^2+p^2\right) \left(-f_H^2 y_L^2-{m_7}^2+p^2\right)} \times \\
 & \frac{1}{\left({m_1}^2 p^2 \left(f_H^2 y_L^2+2 {m_7}^2-2 p^2\right)+\left({m_R}^2-p^2\right) \left({m_7}^2 \left(2 p^2-f_H^2 y_L^2\right)+2 f_H^2 p^2 y_L^2-2 p^4\right)\right)}.
\end{split}
\end{equation}

$\Delta\beta$ is obtained in the same fashion.

\end{widetext}

\bibliography{TadpoleRefs}

\end{document}